\documentclass{ametsocV6.1}

\usepackage{graphicx}
\usepackage{tabularx}

\def\jfm{{\it J.\ Fluid.\ Mech.}}
\def\dsr{{\it Deep-Sea\ Res.}}
\def\dsrI{{\it Deep-Sea\ Res.\ I}}

\def\gfd{{\it Geophys.\ Fluid\ Dyn.}}
\def\jmr{{\it J.\ Mar.\ Res.}}
\def\jas{{\it J.\ Atmos.\ Sci.}}

\def\jpo{{\it J.\ Phys.\ Oceanogr.}}
\def\jm{{\it J.\ Meteor.}}
\def\jgr{{\it J.\ Geophys.\ Res.}}

\def\tagu{{\it Trans.\ Amer.\ Geophys.\ Union} }

\def \BEA {\begin{eqnarray}} 
\def \EEA {\end{eqnarray}}

\def \p {{\textbf p}}
\def \r {{\textbf r}}

\def \K {\textbf{k}}
\def \U {\textbf{U}}

\nolinenumbers

\title{Mesoscale Eddy - Internal Wave Coupling.  III.  The End of the Enstrophy Cascade and Maintenance of Gyre Scale Potential Vorticity Gradients }

\authors{Kurt L. Polzin \aff{a}\correspondingauthor{Kurt L. Polzin, kpolzin@whoi.edu}, Giovanni Dematteis \aff{a}\aff{b}}

\affiliation{\aff{a}{Woods Hole Oceanographic Institution}\\\aff{b}{Universit\`a di Torino, Dipartimento di Fisica}}

\abstract{We assess a prognostic formulation of triple coherence relating to energy exchange between mesoscale eddies and the internal wavefield and compare with observations from the Sargasso Sea.  This effort involves updates to a theory articulated in \cite{M76} that balances eddy induced wavefield perturbations with nonlinearity using a relaxation time scale approximation.  Agreement of the prognostic formulation with data is remarkable and is consistent with eddy-wave coupling dominating the regional internal wave energy budget.  The goodness of this effort reinforces a prior hypothesis that the character of the internal wavefield in the Sargasso Sea is set by this interaction, which, in turn, serves as an amplifier of tertiary energy inputs from larger vertical scales that characterize internal swell.  Extraction of eddy energy happens at the horizontal and vertical scales that characterize baroclinic instability and potential vorticity fluxes.  With this knowledge and confidence, we then speculate on the role that this coupling plays with regards to mesoscale eddy dynamics in the Southern Recirculation Gyre of the Gulf Stream. We argue that this nonlinear relaxation effectively provides a local eddy enstrophy damping consistent with potential vorticity flux observations from the Local Dynamics Experiment.  This happens at spatial scales somewhat smaller than the energy extraction scale and locates the end of the potential enstrophy cascade in the spectral domain as the energy containing scale of the internal wavefield.  The dynamical consequence is that mesoscale eddy - internal wave coupling is responsible for the maintenance of gyre scale potential vorticity gradients.  }

\begin{document}
\nolinenumbers

\maketitle

%
%
%
%
%
%

%

\section{Introduction}
\label{intro}
\subsection{Prologue}
Winds and air-sea exchanges of heat and fresh water are ultimately
responsible for the basin-scale currents, or general circulation of the oceans.  In order to achieve a state where the energy and potential enstrophy (potential vorticity perturbations, squared) of the ocean are not continuously increasing, some form of dissipation is required to balance this forcing.  While the above statement may seem obvious, little is known about how and where this dissipation occurs.

Early theories of the wind driven circulation [\cite{Stommel48}, \cite{Munk50}] view the western boundary as a region where energy and vorticity input by winds in mid-gyre could be dissipated.  Those theories predict Gulf Stream transports that are approximately equal to the interior Sverdrup transport [about 30 Sv, \cite{Schmitz92}] and that are much smaller than observed Gulf Stream transports after the Stream separates from the coast [about 150 Sv, \cite{Johns95}].  Subsequent theories of the wind driven circulation have attempted to address the role of nonlinearity and baroclinicity in increasing Gulf Stream transports above that given by the Sverdrup relation.  

\cite{Hogg83} proposed the existence of two relatively barotropic recirculation gyres on either side of the Stream that combine to increase the total transport.  The recirculation gyres are generally acknowledged to result from potential vorticity fluxes \citep{BOB86,hogg1993toward} generated by the meandering of a baroclinicly unstable Gulf Stream \citep{B82, cronin1997eddyI} and the potential vorticity flux divergence, ultimately relating to an absorption process.  That `absorption' process is uncertain.  

An objective of this study is to develop a quantitative prognostic assessment of energy transfers between mesoscale eddies and the internal wavefield.  However ambitious that might seem, the goal is far grander.  The goal of this study is to articulate the role of internal waves in maintaining gyre scale potential vorticity gradients that are so crucial to Earth system behavior.  Both the objective and the goal are made concrete using prior analyses of data obtained as part of the Local Dynamics Experiment (LDE) of the PolyMode III program executed in 1978-1979 and located at the south-western edge of the Southern Recirculation Gyre.  Once this has been accomplished, we can assess such things as whether numerical simulations realistically capture the 'absorption' process referred to above.

\subsection{Goal}
The goal of this study is encapsulated with three distinct results.  The first is the documentation by \cite{BOB86} of statistically significant potential vorticity fluxes using the LDE array data, Section \ref{budgets}.\ref{Enstrophy_Budget}.  These are directed to the SSW.  The second is the documentation of background potential vorticity gradients by \cite{Robbins} \footnote{See also \cite{wijffels2024mesoscale} for a gridded climatology of potential vorticity fields}.  At the level of the \cite{BOB86} potential vorticity flux estimates, the potential vorticity gradients are essentially zonal and indistinguishable from variations in the planetary rotation rate.  The third result is a simple consequence of the prior two:  potential vorticity fluxes directed across potential vorticity contours imply potential enstrophy production.  In order to balance this production term, one has three options:  time dependence of potential enstrophy, dissipation, or nonlocal transports (triple correlations).  We regard the first as unlikely.  Consequences of the third are discussed in \cite{RY82}:  in the limit of small interior mixing, the role of triple correlations is to transport enstrophy to the boundaries, where it is dissipated.  As a consequence, potential vorticity in the gyre interior is homogenized.  Since this is inconsistent with the observations of \cite{Robbins}, we regard a balance between enstrophy production and enstrophy dissipation as far more likely.  Basic scaling of the quasi-geostrophic potential vorticity equation directs us to a viscous dissipation process related to the characterization of energy exchanges between internal waves and mesoscale eddies \citep{polzin2010mesoscale, Bow} rather than diapycnal mixing as the dominant dissipation mechanism.  Motivated by the foundational work in \cite{Stommel48} and \cite{Munk50} concerning barotropic vorticity models, the goal of this work is to frame an answer for, `what is friction?' in the context of a rotating stratified fluid that speaks to gyre scale vorticity dynamics.  The logical consequence is a highly non-dualistic proposition that internal waves are fundamental to maintaining the gyre scale potential vorticity gradients which are key to Earth system behavior.  

\subsection{Objectives}
The objectives of this study are (i) to provide a quantitative prognostic assessment of energy transfers between the background internal wavefield and the mesoscale eddy field that are the result of the mesoscale inducing momentum flux anomalies in the background wavefield, (ii) to compare those prognostic estimates with field data from the Local Dynamics Experiment reported in \cite{polzin2010mesoscale} and (iii) to sketch a physically plausible characterization of enstrophy dissipation in which we ground the effort with potential vorticity fluxes and enstrophy production estimates using data from this same experiment \citep{BOB86}.  This last effort leads us to a plausible set of physics that defines the vertical wavenumber bandwidth of the regional internal wave spectrum.    

Intuitive understanding about our pursuit of these objectives can  be attained by leaning on the turn of phrase that energy exchange can be accomplished if either the forcing is in phase with the damping or if the forcing is in resonance.  Here one can consider forcing as wave refraction projecting onto excursions in wavenumber space through ray tracing and damping as the result of wave-wave interactions.  Both issues need considerable development.  The issue of resonance is also relevant given the interpretation of Tom Sanford's original MODE data set \citep{Sanford75,L76} as an example of a phase velocity - group velocity resonance in \cite{polzin2008mesoscale}.  We are not yet sufficiently sophisticated to deal with this as there are profound differences between amplitude modulated (AM) and frequency modulated (FM) systems in key concepts such as resonant bandwidth at finite amplitude \citep{polzin2017oceanic,lvov2024generalized,polzin2025one}.  While wave turbulence and the bulk of nonlinear oscillator interpretations concern AM systems, ray tracing is an FM system and this leads to an {\em ad hoc} mash up.  

A prognostic model for the energy exchange is built by modifying \cite{M76}'s presentation of a radiation balance scheme \citep{MO} that has broad parallels with Boltzmann's equation for a rarefied gas.  We encourage the reader to revisit their upper level undergraduate textbooks concerning:  (1) A derivation of Boltzmann's equation.  Here number density in physical space is replaced by wave action spectral density.  Pseudomomentum, wavenumber times action spectral density, replaces momentum and the particle momentum coordinates of Boltzmann's equation are replaced by wavenumber.  Relaxation to thermal equilibrium through particle scattering will be replaced by relaxation to a stationary state through nonlinear wave-wave interactions, {\it aka Wave Turbulence} \citep{zakharov2012kolmogorov,nazarenko2011wave}.  Rather than collisional cross sections representing particle interactions, energy and momentum are transferred between three waves along a resonant manifold.  (2) The Quantum Mechanics textbook contains similar relevant material about Open Quantum Systems in which the Heisenberg notation corresponds to ray coordinates and the currency concerns Hamiltonians that will motivate our development of Boltzmann's analog.  It will also remind the reader about subtleties concerning waves (plane waves) and particles (wave packets).  Here the working basis for plane waves is the amplitude modulated theory of wave turbulence, distinct from the frequency modulated basis for a wave packet.  (3) The Classical Mechanics textbook will provide grounding with Hamiltonians, generalized coordinates and Louiville's equation.  (4) A graduate level text concerning the structure of a formal WKB expansion will be useful when discussing the issue of vorticity dynamics.  

\subsection{Outline}
This paper is constructed as follows.  In Section \ref{Prognostic}.\ref{EnergyEq} we frame internal wave energy exchanges with the mesoscale in terms of spectral moments of the internal wavefield.  In Section \ref{Prognostic}.\ref{Boltzman} we develop the wave action Boltzmann analog as a prognostic model for those spectral moments.  Section \ref{Prognostic}.\ref{Forcing} attempts to build an intuitive understanding of asymptotic system behavior for `forcing' in the extreme scale separated limit of ray tracing.  Section \ref{Prognostic}.\ref{Damping} presents the issue of damping.  This effort provides an alternative derivation to that presented in \cite{M76}.  Section \ref{Coupling}.\ref{navigation} navigates the algebra behind the elements of the preceding two sections, synthesizing the  Energy and Boltzmann equations.  Section \ref{Coupling}.\ref{RegionalSpectrum} defines the regional internal wave spectrum $n^{(0)}(\p)$.  Propagation time scales $\tau_p$ are addressed in Section \ref{Coupling}.\ref{PropagationTimeScales}.  Numerical evaluations and a comparison with the LDE based estimates of energy transfer from \cite{polzin2010mesoscale} are presented in section \ref{Coupling}.\ref{Evaluations}.  Regional energy and potential enstrophy budgets are discussed in Section \ref{budgets}\ref{IWenergy} and \ref{budgets}.\ref{Potential_Vorticity}, respectively.  The goal of this work is addressed in Section \ref{budgets}.\ref{Potential_Enstrophy}.\ref{ScaleDependent} where transfers of eddy potential enstrophy to packet scale wave momentum are estimated using a length scale dependent closure.  We summarize our results in Section \ref{Summary}.\ref{Results} and attempt to frame both future work and potential implications for Earth system behavior in Sections \ref{Summary}.\ref{Optics} and \ref{Summary}.\ref{Interpretation}.  Appendix A describes alternative efforts at assessing and formulating internal wave - mesoscale eddy coupling.  Appendix B addresses uncertainty metrics and parameter sensitivity.

\section{Kinetic Theory for the Ocean}\label{Prognostic}
\subsection{Energy equation}\label{EnergyEq}
We start by presenting an equation for the perturbation energy $E = E_k + E_p$:
\begin{eqnarray}\label{Energy1}
(\frac{\partial}{\partial t} + U \cdot \nabla_h ) (E_k + E_p) &+& \nabla \cdot \overline{ \pi \bf{ u} } \;\;+\;\; \mathcal{N}\mathcal{L} \\ 
& = & -\overline{ u u } U_x 
-\overline{ u v } U_y 
-\overline{ v u } V_x
-\overline{ v v } V_y 
-\overline{ u w } U_z
- \overline{b u} ~ B_x /B_z
-\overline{ v w } V_z
-\overline{b v} ~ B_y /B_z \nonumber
\end{eqnarray} 
in which the velocity ($u,v,w$), density $\rho$ $(b=-g\rho/\rho_0)$ and pressure $p$ $(\pi=p/\rho_0)$ fields have been represented using lower case for the internal wave perturbation and upper case for the background.  Nonlinear internal wave contributions are schematically rendered as $\mathcal{NL}$.  

We structure the conversation by characterizing the background flow using
\begin{eqnarray}
S_n & \equiv & U_x - V_y \nonumber \\
\Delta & \equiv & U_x + V_y \nonumber \\
S_s & \equiv & V_x + U_y \nonumber \\
\zeta & \equiv & V_x - U_y \nonumber \\
fU_z & \equiv & B_y + \delta_y \nonumber \\
fV_z & \equiv & -B_x -\delta_x 
\label{eq:BackgroundDefinitions}
\end{eqnarray}
and explicitly acknowledge ageostrophic departures from the thermal wind relation $(B_x,B_y)=f(-V_z,Uz)$.  The variables $S_n$ and $S_s$ are components of the horizontal deformation rate of strain tensor, distinct from the vertical component of relative vorticity $\zeta$ and horizontal divergence $\Delta$.  A geometric interpretation of energy exchange as wave momentum (stress) - eddy strain relations is pursued in Section \ref{Prognostic}.\ref{Forcing}.  

Thus, 
\begin{eqnarray}\label{Energy2}
(\frac{\partial}{\partial t} + U \cdot \nabla_h ) (E_k + E_p) &+& \nabla \cdot \overline{ \pi \bf{ u} } \;\; + \;\; \mathcal{N}\mathcal{L}\nonumber \\ 
\nonumber \\
& = & -\overline{ u v } S_s -(\overline{uu} -\overline{ v v } )S_n/2 - (\overline{uw} - \frac{f}{B_z}\overline{bv})U_z -(\overline{vw} + \frac{f}{B_z} \overline{bu} )V_z \nonumber \\ && \;\;\;\;\;\;\;\;\;\;\; -(\overline{uu}  + \overline{ v v }) \Delta/2  +\overline{bv}\frac{\delta_y}{B_z} + \overline{bu}\frac{\delta_x}{B_z}
\end{eqnarray}
Note the absence of an explicit reference to the relative vorticity $\zeta$ in this energy equation.  We then represent the momentum and buoyancy fluxes in terms of energy density using polarization relations \cite{MO,polzin2011toward}.  
Assuming a plane wave formulation of $a \exp^{i[{\bf r}\cdot {\bf p}-\sigma t]}$, the linearized f-plane equations of motion can be manipulated to provide: 
\begin{eqnarray}\label{eq:PolarizationRelations}
u & = & \frac{(k+ifl/\omega)}{k_h} ~ a ~ {\rm e }^{i({\bf p \cdot r} -\sigma t)}\label{PR1}
\nonumber \\
v & = & \frac{(l-ifk/\omega)}{k_h} ~ a ~ {\rm e }^{i({\bf p \cdot r} -\sigma t)}
\nonumber\\
w & = &  -\frac{k_h}{m} ~ a ~ {\rm e }^{i({\bf p \cdot r} -\sigma t)} \\
b & = &  \frac{ik_h N^2}{\omega m} ~ a ~ {\rm e }^{i({\bf p \cdot r} -\sigma t) } \nonumber 
\end{eqnarray}
We use the hydrostatic versions and an intrinsic frequency $\omega = \sigma - \p \cdot \U$ in place of $\sigma$.  The prefactor in these polarization relations is such that the wave amplitude $a$ is normalized to represent the sum of horizontal kinetic $E_k$ and potential $E_p $:
\begin{equation}
E_k + E_p = aa^{\ast} ~. \nonumber
\end{equation}
These polarization relations invoke only the Doppler shift and are the lowest order of a scale separation in ray tracing.  
The polarization relations are then introduced into the energy equation using $\overline{xy}=xy^{\ast}$:  
\begin{eqnarray}\label{Energy3}
(\frac{\partial}{\partial t} + U \cdot \nabla_h ) aa^{\ast} &+& \nabla \cdot \overline{ \pi \bf{ u} } \;\; + \;\; \mathcal{N}\mathcal{L} \\ 
\nonumber \\
& = & \frac{aa^{\ast}}{\omega} \big[ \frac{\omega^2-f^2}{\omega} \big( - \frac{kl}{k_h^2}  S_s - \frac{k^2-l^2}{k_h^2} S_n/2 - \frac{k}{m} U_z - \frac{l}{m} V_z - \Delta/2 \big)  + \frac{f}{\omega} \big(\frac{k}{m} \delta_y + \frac{l}{m} \delta_x \big)\big] \; .\nonumber
\end{eqnarray}
The next step is to develop closures for these energy transfer terms as statistical moments of the action density $aa^{\ast}/\omega$.    

\subsection{The Wave Action Boltzmann Analog }\label{Boltzman}

At this juncture we, revisit \cite{M76} in which statistical closures for action weighted ensemble moments such as $\langle kl \rangle aa^{\ast}/\omega$ and $\langle k^2-l^2 \rangle aa^{\ast}/\omega$ are presented.  These moments capture the distortion of a spatially homogeneous, isotropic background internal wave spectrum by a background flow and damping of those perturbations through a nonlinear relaxation process, resulting in permanent energy transfers.  The result is nearly that of \cite{M76}.  However, \cite{M76}'s presentation, which invokes a perturbation expansion and variational calculus, can be substantially simplified.  The essential result requires little more than algebra.  We present that algebra within this Section to demystify the rest of the paper.  

At the outset, the intent is to characterize interactions between the background internal wave spectrum and the mesoscale eddy field.  The underlying description for this background is the parametric spectral representation introduced in \cite{GM72} with vertical wavenumber bandwidths of $4 \leq j_{\ast} \leq 20$ that vary on a regional basis \citep{polzin2011toward}.  We are {\it not} focused upon low mode ($1 \leq j \leq 2$) near inertial waves or low mode internal tides.  We are focused on the analog of wind waves rather than swell in the surface wavefield.  With this, we are safely within an extreme scale separated paradigm and the analysis is grounded in action conservation and ray tracing concepts.  The regional spectrum is presented in Section \ref{Model}.\ref{RegionalSpectrum} . 

For a single wave packet, the presumption of scale separated interactions implies that wave action spectral density $n(\p(\r(t)))$ is conserved along wave characteristics in space $\r(t)$ and wavenumber $\p(\r)$:
\begin{equation}
\frac{\partial n_3(\p(\r(t)))}{\partial t} + \dot{\r} \cdot \nabla_{\r} n_3(\p(\r(t))) + \dot{\p} \cdot \nabla_{\p} n_3(\p(\r(t))) = 0
\label{eq:ActionConservation}
\end{equation}
in which $\p = (k, l, m)$ is the 3-D wavevector with $\K=(k, l)$ and $k_h=(k^2+l^2)^{1/2}$.  We employ subscripts to denote the dimensionality of the spectral density.  We will reference a horizontally isotropic spectral density and move between 2-D and 3-D using $n_2(k_h,m)=k_h n_3(k_h,m)$.  Furthermore, wave action spectral density is related to energy density as $n(\p) = E(\p)/\omega$ and the change of variables between frequency and wavenumber space utilizes $n_2(k_h,m) dk_h = n_2(\omega,m) d\omega $.  The factor $\dot \r$ represents the rate of refraction in physical space $\r$ (advection plus group velocity) and $\dot \p$ the rate of refraction in the spectral domain.  These are local functions of space, time and wavenumber.  We will represent the background variables using upper case notation.  Horizontal velocities are $\U=(U,V)$ and the background wavevector is $(K, L, M)$.  We direct the reader to \cite{lvov2024generalized} for an actual derivation of (\ref{eq:ActionConservation}).  

We will work with an Eulerian representation by averaging over all wave trajectories that bring wave packets to position $\p$ at $\r$ and $t$, $n(\p; \r, t)$:
\begin{equation}
\frac{\partial n(\p; \r, t)}{\partial t} + \dot{\r} \cdot \nabla_{\r} n(\p; \r, t) + \dot{\p} \cdot \nabla_{\p} n(\p; \r, t) = S_o - S_i + \mathcal{N}\mathcal{L}(n(\p; \r, t)) + O(\frac{\ell}{\mathcal{L}}). 
\label{eq:RadiationBalance}
\end{equation}
in which nonconservative processes are represented on the right-hand-side.  Sources such as near-inertial wave generation from wind forcing or internal tide generation from sloping topography are represented by $S_o$, sinks such as wave-breaking by $S_i$.  Nonlinear interactions, which conserve energy and pseudomomentum amongst three waves rather than the wave action, Section \ref{Prognostic}.\ref{Damping}, are represented schematically as $\mathcal{ NL}$.  We direct the reader to \cite{MO} for a presentation of this radiation balance scheme and to \cite{eden2019numerical, dematteis2021downscale, dematteis2022origins, lvov2024generalized, polzin2025one, dematteis2024interacting} for recent work on the nonlinear transfers.  In order to arrive at the action conservation statement (\ref{eq:ActionConservation}), there is a discard of terms representing the packet structure having a spatial scale $\mathcal{L}$, assumed to be large relative to the inverse wavenumber $\ell$ \citep{gershgorin2009canonical}, see also \cite{lvov2024generalized}.  The envelope structure of a wave packet sets a pseudomomentum flux divergence.  If the envelope structure is finite in both horizontal dimensions, the curl of the pseudomomentum flux divergence is nonzero, and the curl of this divergence represents a PV anomaly \citep{BM05}.  This is a key issue as concerns the role of internal waves in the mesoscale eddy potential enstrophy budget (Section \ref{budgets}.\ref{Enstrophy_Budget}) and a notable departure from \cite{M76}, in which the role of such residual circulations was not realized.  We do not delve into this in detail, simply pointing out that the quantitative connection appears in \cite[e.g.][]{bretherton1969mean,BM05,xie2015generalised,wagner2015available}.  

We assume the internal wavefield can be represented as the sum of a quasi-stationary, quasi-homogeneous isotropic component $n^{(0)}(\p)$ and two that vary in space and time in response to eddy interactions:
\begin{equation}
n_3 = n_3^{(0)}(\p) + n_3^{(1)}(\p; \r,t) + n_3^{(2)}(\p; \r,t) 
\label{eq:Decomposition}
\end{equation}
where $n_{3}^{(1)} \ll n_{3}^{(0)}$ and $n_{3}^{(2)} \ll n_{3}^{(0)}$.  The $n^{(0)}$ balance is attained as sources $S_o$ and sinks $S_i$ being connected by nonlinear interactions $\mathcal{N}\mathcal{L}$.  The phrases 'quasi-stationary' and 'quasi-homogeneous' imply a multiple time scale process that becomes concrete within the context of the regional and seasonal variability of the internal wavefield documented in \cite{polzin2011toward}.  The intent of $n^{(1)}$ is to capture the modulation of ensemble averaged wave action by the eddies in the extreme scale separated limit assumed in (4).  The intent of $n^{(2)}$ is to address $O(\ell/{\mathcal L})$ factors dropped from the polarization relations (4), the eikonal relations $\dot{\bf p}$, group velocities within $\dot{\bf r}$ and, importantly, residual circulations on the packet scale.  Motivated by our understanding of the extreme scale separated nature of the energy exchange process, we neglect $n^{(2)}$ in an unconstrained approximation.  We revisit this issue in Section \ref{budgets} when we pick up the issue of the potential enstrophy budget and in Section \ref{Summary}.\ref{Optics}.\ref{WKB}.  Here we proceed by substituting the decomposition (\ref{eq:Decomposition}) and include $\dot \p$ (Section \ref{Prognostic}.\ref{Forcing}) associated with coupling to the mesoscale.  Subtracting the homogeneous $n^{(0)}$ balance from (\ref{eq:RadiationBalance}) provides
\begin{equation}
\frac{\partial n^{(1)}}{\partial t} + \dot{\r} \cdot \nabla_{\r} \; n^{(1)} + \dot{\p} \cdot \nabla_{\p} [n^{(0)} + n^{(1)}] \cong \mathcal{N}\mathcal{L}(n^{(1)}; n^{(0)}) . 
\label{eq:SpaceTimeRadiationBalance}
\end{equation}
We cast $\mathcal{N}\mathcal{L}(n^{(1)}(\p; \r, t); n^{(0)}(\p))$ as a relaxation time scale (Sections \ref{Prognostic}.\ref{Damping},  
\begin{equation}
\mathcal{N}\mathcal{L}(n^{(1)}; n^{(0)}) = \tau_r^{-1}(n^{(0)}) \; n^{(1)} \; . \nonumber
\end{equation}
Assuming that 
\begin{equation}
n^{(1)}(\p; \r, t) \ll n^{(0)}(\p)\; , \nonumber
\end{equation}
we rearrange (\ref{eq:SpaceTimeRadiationBalance}) to obtain
\begin{equation}
\frac{\partial \; n^{(1)}}{\partial t} + \dot{\r} \cdot \nabla_{\r} \; n^{(1)} - \tau_r^{-1}(n^{(0)}) \; n^{(1)} = - \dot{\p} \cdot \nabla_{\p} \; n^{(0)} . 
\label{eq:NearlyThere}
\end{equation}
In \cite{M76}, $n^{(0)}, n^{(1)} \; {\rm and} \; n^{(2)}$ are assumed to be an ordered expansion in terms of a small parameter relating advection to the wave phase speed, and thus $\dot{ \r} = {\bf U} + {\bf C}_g \cong {\bf C}_g$ in (\ref{eq:NearlyThere}).  Non-local effects in (\ref{eq:NearlyThere}) are addressed by identifying the space-time dependence of $n^{(1)}$ as having the background Fourier components, i.e. $n^{(1)} \propto e^{i[Kx+Ly+Mz-\Omega t]}$ and Fourier transforming (\ref{eq:NearlyThere}).  
\begin{eqnarray}\label{BadBoy}
i(K C_g^x + L C_g^y + MC_g^z - \Omega) - \tau_r^{-1}(n^{(0)}) \; n^{(1)} &=& - \dot{\p} \cdot \nabla_{\p} \; n^{(0)} \nonumber \\
n^{(1)} &=& \frac{-1}{i(K C_g^x + L C_g^y + MC_g^z - \Omega) - \tau_r^{-1}(n^{(0)})} \dot{\p} \cdot \nabla_{\p} \; n^{(0)} \nonumber \\
n^{(1)} &=& \frac{-1}{i\tau_p^{-1} - \tau_r^{-1}} \dot{\p} \cdot \nabla_{\p} \; n^{(0)} 
\end{eqnarray}
in which $\tau_p$ represents a propagation time scale:  
\begin{equation}
\tau_p^{-1} = K C_g^x + L C_g^y + MC_g^z - \Omega \; .
\label{PropagationTimeScale}
\end{equation}
The mesoscale velocity gradients are then assumed, for the sake of simplicity, to be concentrated at a single scale.  Energy exchange occurs between the mesoscale and the internal wavefield occurs as the damping is in phase with the forcing, and we identify this as the real part of $n^{(1)}(\p; \r, t)$, i.e.  
\begin{equation}
\Re (n^{(1)}(\p; \r, t)) = - \frac{\tau_r}{1+(\tau_r / \tau_p)^{2}} \; \dot \p \cdot \nabla_{\p} \; n^{(0)}(\p) \; .
\label{eq:HolyGobSmack}
\end{equation}
in which $\Re$ means the real part.  

Having solved for $n^{(1)}(\p; \r, t)$, estimates of the energy transfer are obtained from (\ref{Energy3}) by identifying $aa^{\ast}/\omega$ in terms of the action spectral density $n^{(1)}(\p; \r, t)$ and simply integrating over wavenumber.  In the vertical coordinate, 
\begin{equation}
\overline{uw} -\frac{f}{N^2}\overline{vb} = \int d\p \; k \; C_g^z \; n^{(1)}(\p; r, t)
\label{VerticalMomentum}
\end{equation}
which in turn gives the energy transfer (\ref{Energy3}) with the mesoscale, $[ \overline{uw} - \frac{f}{N^2} \overline{vb} ]\; U_z$, facilitating identification of a vertical viscosity,  $[ \overline{uw} - \frac{f}{N^2} \overline{vb} \; ] = -(\nu_v + \frac{f^2}{N^2}K_h ) U_z$.  

\begin{quote}
    Despite strategic differences concerning the definitions of $n^{(1)}$ and  $n^{(2)}$, this is the essence of \cite{M76}.  The formulation allows one to clearly identify the roles of forcing, damping and resonance in energy exchange.  That assessment of energy exchange at wavenumber $\p$ reduces to an intensive algebraic exercise.  
\end{quote}

In our approach, the reasoning is more subtle.  We note that the left-hand-side of (\ref{eq:RadiationBalance}) applies to a single realization of $\dot \r$ and $\dot \p$.  As detailed in Section \ref{Prognostic}.\ref{Forcing} below, the eddy forcing in (\ref{eq:NearlyThere}), i.e. the right-hand-side, is proportional to the deformation rate of strain.  The deformation rate of strain, in turn, is independent of the sign of the velocity, such that in executing an ensemble average of (\ref{eq:NearlyThere}), $\dot{ \r} = {\bf U} + {\bf C}_g \cong {\bf C}_g$ and we wind up at the same place, (\ref{eq:HolyGobSmack}).  

Having presented both an energy equation and a Boltzmann analog, we step back to develop intuitive notions of what constitutes `forcing' and `damping' in the following subsections.  

\subsection{Forcing}\label{Forcing}


 
In this Boltzmann analogy, $\dot \p$ plays the role of forcing.  This follows from Newton's $force \; = \;mass \; \times \; acceleration$, in which $mass \; \times \; acceleration $ is the time rate of change of $momentum$.  Here, pseudomomentum ${\bf \p}  n_3({\bf \p}(\r(t)))$ replaces $momentum$.  Since the action spectral density $n_3(\p(\r(t))$ is conserved following a ray path, the time rate of change of pseudomomentum is given by $\dot \p$.  

Our interest is in small amplitude waves propagating in a larger amplitude, larger {\em horizontal} scale background, for which we refer the reader to \cite{lvov2024generalized} for an actual derivation of action spectral density conservation.  Wave action spectral density 
\begin{equation}
    n(\p(\r)) = a(\p(\r))a^{\ast}(\p(\r))/\omega
\end{equation}
is conserved  
\begin{eqnarray}\label{TransportEquation}
\frac{\partial n({\p(\r)})}{\partial t}+ {\dot \r}\cdot 
\nabla_{\r} \; n(\p(\r)) + {\dot \p} \cdot\nabla_{\p}\; n(\p(\r) ) =0,\nonumber
\end{eqnarray}
along characteristics (ray trajectories) defined by 
\begin{eqnarray}\label{Characteristics.org}
\dot \r(t) & \equiv & \nabla_\p \sigma_{\p,\r} = {\bf U}+{\bf C_g},
\\ 
\dot \p(t) & \equiv & - \nabla_\r\sigma_{\p,\r} \; ,  \nonumber
\end{eqnarray}
in which the intrinsic frequency $\omega = \sigma - {\bf p} \cdot {\bf U }$ plays the role of the Eulerian frequency $\sigma$ in the polarization relations.  Consistent with the polarization relations (\ref{eq:PolarizationRelations}), we use only the Doppler shift for $\dot \p(t)$:  
\begin{eqnarray}\label{eq:Eikonal}
    \dot{k} = -kU_x - lV_x - mW_x \nonumber\\
    \dot{l} = -kU_y - lV_y - mW_y \\
    \dot{m} = -kU_z - lV_z - mW_z \nonumber
\end{eqnarray}
where we neglect vertical gradients of background stratification.  In combination, (\ref{eq:PolarizationRelations}) and (\ref{eq:Eikonal}) represent the lowest order description of extreme scale separated internal wave - mean flow interactions.  

The LDE moored array provides thermocline level estimates of relative vorticity and vertical shear at periods greater than one day that are small, $[R_o, \; F_r]=[(V_x-U_y)_{{\rm rms}}/f,\; (U_z^2+V_z^2)^{1/2}/N] = [0.025, \; 0.031]$.  Thus terms in (\ref{eq:Eikonal}) involving the $O(R_o^2)$ vertical velocity are small.  

If the background $n^{(0)}$ is horizontally isotropic, then the horizontal components of the forcing become
\begin{eqnarray}
\dot{\p}{\bf \cdot} \nabla_{\p} n^{(0)}(\p) & \rightarrow & \big[ -k\big( \frac{S_n+\Delta}{2} \big)-l\big( \frac{S_s+\zeta}{2} \big) ; -k\big( \frac{S_s -\zeta}{2} \big)-l\big( \frac{\Delta - S_n}{2} \big) \big] \cdot  \nabla_{(k;l)} n^{(0)}  \nonumber \\
& = & \big[ -klS_s - (k^2-l^2) S_n/2 - (k^2+l^2)\Delta / 2 \big] k_h^{-1} \nabla_{k_h} n^{(0)} 
\label{SourceTerm}
\end{eqnarray}
where we have inserted the deformation strain - relative vorticity - horizontal divergence definitions in (\ref{eq:BackgroundDefinitions}) and then invoked the chain rule.  
Energy exchange (\ref{Energy3}) occurs as the wave stresses correlate with the components of the deformation rate of strain tensor, i.e. 
\begin{eqnarray}\label{HorizontalPseudomomentum}
-\overline{ u u } \; U_x -\overline{ v v } \; V_y = &
-\frac{(k^2-l^2)N^2}{k_h^2N^2+f^2m^2} \; n^{(1)} \; S_n/2 \nonumber \\ 
-\overline{ u v } \; U_y -\overline{ v u } \; V_x = &
-\frac{kl N^2}{k_h^2N^2+f^2m^2} \; n^{(1)} \; S_s \; .
\end{eqnarray}  
Equations (\ref{HorizontalPseudomomentum}) represent the energy exchange for a single wave in which there is a clear geometric interpretation of wavenumber stretching, rotation and dilation that parallels that of particle pair separation \citep{okubo1970horizontal, hua1998exact} in the limit that the group velocity is much smaller than background advection \citep{jones1969ray} \cite{BM05}.  The sign of the energy exchange in this limit then relates to how the wave particle velocities expressed in the polarization relations (4) are aligned with or against the components of the deformation rate of strain tensor, $S_n$ and $S_s$ tensor \citep{polzin2010mesoscale, marshall2012framework}.  
%
%
%

Assuming that the time dependence of horizontal wavenumber exceeds that of the background gradients along a ray, upon differentiation with respect to time of the eikonal equations (\ref{eq:Eikonal}) one \citep{jones1969ray,BM05} arrives at 
\begin{equation}\label{OW}
    [\ddot{k},\ddot{l}] = [k,l] (S_n^2+S_s^2-\zeta^2) \; .
\end{equation}
which implies exponential growth and decay of horizontal wavenumber along the ray if the rate of strain variance dominates the relative vorticity variance and oscillatory solutions otherwise.  From casual inspection of the background gradients in the forcing term (\ref{SourceTerm}), the relation of forcing to $n_3^{(1)}$ in (\ref{eq:HolyGobSmack}) and those gradients involved in the energy transfer (\ref{HorizontalPseudomomentum}), we conclude that the issue of relative vorticity $\zeta$ is a red herring in the energetics of extreme scale separated interactions. 

In the vertical coordinate, if the background is in thermal wind balance, 
\begin{eqnarray}\label{eq:verticalEnergetics}
 -\overline{ u w }\;  U_z -N^{-2} \; \overline{b v}\; B_y & = kC_g^z \frac{aa^{\ast}}{\omega}U_z \\
-\overline{ v w } \; V_z -N^{-2} \overline{b u} \; B_x\nonumber & = l C_g^z \frac{aa^{\ast}}{\omega} V_z 
\end{eqnarray}
and $(k,l) C_g^z aa^{\ast}/\omega$ can be parsed as the vertical flux of horizontal angular momentum across an isopycnal surface \citep{jones1967propagation,bretherton1969momentum} which has a rich history in the zonal mean literature \citep{AHL,polzin2010mesoscale} under the phrase Eliassen-Palm flux \citep{EP}.  In contrast to that parallel shear flow paradigm, which features linear growth of vertical wavenumber in time as the horizontal wavenumber aligns with the background flow, here the vertical wavenumber is slaved to the horizontal wavenumber via $\dot{m} = -kU_z -lV_z$ and thus asymptotically experiences similar exponential growth and decay \citep{jones1969ray, BM05}.   

While one might quibble regarding the amount of time the deformation rate of strain can be considered constant along the ray, one has arrived at an intuitive understanding why horizontal stress and horizontal strain can be correlated.  See Figures 3-4 and Figures 5-6 of \cite{polzin2010mesoscale} for observations documenting the horizontal stress-strain and vertical stress-shear relations.  

If we appealed to the small advection limit envisioned by M\"uller and assumed that the mesoscale is time invariant, system behavior is significantly different \citep{kafiabad2019diffusion}:  horizontal wavenumber magnitude is constant and horizontal wavenumber azimuth is diffused by the vertical vorticity $\zeta$.  There is no energy exchange at leading order in this limit \citep{dong2020frequency}.  

\subsection{Damping}\label{Damping}
Our intuition leads us to specifying the relaxation timescale $\tau_R$ as a return to the regional stationary state $n^{(0)}(\p)$.  This stationary state is defined as a balance between sources, sinks and nonlinearity, (\ref{eq:RadiationBalance}) and (\ref{eq:Decomposition}).  While the regional character of $n^{(0)}$ stems at least in part from wave-mean interactions, the $n^{(0)}$ definition provided in Section \ref{Model}.\ref{RegionalSpectrum} is empirical rather than prognostic.  

Here there is an analogy to Boltzmann's equation, in which the relaxation time scale is conceived as a scattering process and the tendency is toward thermodynamic equilibrium.  The relaxation process here comes from Wave Turbulence, in which the integrand of the associated kinetic equation describing resonant interactions between three waves takes the place of collisional cross sections and the evaluation takes place on a resonant manifold.  Assessment and evaluation of these is a complicated affair.  

The Wave Turbulence kinetic equation
\begin{eqnarray}
\frac{\partial}{\partial t} n_{\p} =  \displaystyle 
4\pi\int d\p_1 d\p_2 & \Big( \mid V_{\p_1,\p_2}^{\p} \mid^2
\delta(\p-\p_1-\p_2)\delta(\sigma-\sigma_{1}-\sigma_{2})
      [n_{\p_1}n_{\p_2}-n_{\p}[n_{\p_1}+n_{\p_2}]]
      \nonumber
      \\ & \left.- \mid
      V_{\p_2,\p}^{\p_1} \mid^2 \delta(\p-\p_1+\p_2)
      \delta(\sigma-\sigma_{1}+\sigma_{2})
      [n_{\p_2}n_{\p}-n_{\p_1}[n_{\p_2}+n_{\p}]]
      \right.
      \nonumber
      \\
            & - \mid
      V_{\p,\p_1}^{\p_2} \mid^2 \delta(\p+\p_1-\p_2)\delta(\sigma+\sigma_{1}-\sigma_{2})
      [n_{\p}n_{\p_1}-n_{\p_2}[n_{\p}+n_{\p_1}]] \Big) .  \nonumber \\
\label{BroadenedKineticEquation}
\end{eqnarray}
  The resonant manifold is defined by the three wavenumbers ${\bm p},\; {\bm p}_1$ and ${\bm p}_2$ and associated frequencies $\sigma$, $\sigma_1$, $\sigma_2$ fulfilling the resonant conditions:  
\begin{eqnarray}\label{ResonantManifold}
\sigma & = & \sigma_1 + \sigma_2 ; \;\;\;\;\; \;\;\;\;\;\; \bf{p} = \bf{p}_1 + \bf{p}_2  \nonumber \\
\sigma & = &  \sigma_1 - \sigma_2 ; \;\;\;\;\;\;\;\; \bf{p} =  \bf{p}_1-  \bf{p}_2   \nonumber \\
\sigma & = &  \sigma_2 - \sigma_1 ; \;\;\;\;\;\; \bf{p} = \bf{p}_2-  \bf{p}_1 \; .  \
\end{eqnarray}
The kinetic equation (\ref{BroadenedKineticEquation}) is a six-dimensional integral for a spatially homogeneous action spectral density tendency at wavenumber $\p$.  The integrand represents the analog of collisional cross-sections $\mid V_{\p_1,\p_2}^{\p} \mid^2$ (\ref{eq:CollisionalCrossSections}) along a resonant manifold, (\ref{ResonantManifold}), that is defined as three wave solutions for wavenumbers constrained by a dispersion relation $\sigma(\p)$.  This is an amplitude modulated theory in which $n(\p)$ is independent of position $\r$ and thus has a Doppler shift defect \citep{holloway1982interaction,polzin2017oceanic,polzin2025one} in an extreme scale separated limit that we shall navigate in our choices of relaxation time scales.  The cross-sections are complicated algebraic expressions:  

\begin{align}
\mid V_{{\bm p}_1,{\bm p}_2}^{{\bm p}} \mid^2 = 
\frac{N^2}{32 g}
\left(
\left(
\frac{k {\bf k}_1 \cdot {\bf k}_2}{k_1 k_2} \sqrt{\frac{\sigma_1 \sigma_2}{\sigma}}
+ \frac{k_1 {\bf k}_2 \cdot {\bf k}}{k_2 k} \sqrt{\frac{\sigma_2 \sigma}{\sigma_1}}
+ \frac{k_2 {\bf k} \cdot {\bf k}_1}{k k_1} \sqrt{\frac{\sigma \sigma_1}{\sigma_2}}
\right.
\right.
\nonumber\\
\left.
\left.
+ \frac{f^2}{\sqrt{\sigma \sigma_1 \sigma_2}}
\frac{k^2 {\bf k}_1 \cdot {\bf k}_2 - k_1^2 {\bf k} \cdot {\bf k}_2 - k_2^2 {\bf k} \cdot {\bf k}_1}{k k_1 k_2}
\right)^2
\right.
\nonumber\\
\left.
+
\left(
f \frac{{\bf k}_1 \cdot {\bf k}_2^{\perp}}{k k_1 k_2}
 \left(\sqrt{\frac{\sigma}{\sigma_1 \sigma_2}} (k_1^2 - k_2^2)
 - \sqrt{\frac{\sigma_1}{\sigma_2 \sigma}}  (k_2^2-k^2)
 - \sqrt{\frac{\sigma_2}{\sigma \sigma_1}} (k^2-k_1^2)\right)
\right)^2
\right)~\ , \label{eq:CollisionalCrossSections}
\end{align}
in which ${\bf k}_2^{\perp}=(-y,x)$ for horizontal wavenumber ${\bf k} = (x,y)$ and $k=\mid {\bm k} \mid$.  

\begin{figure}
\includegraphics[width=1.0\textwidth]{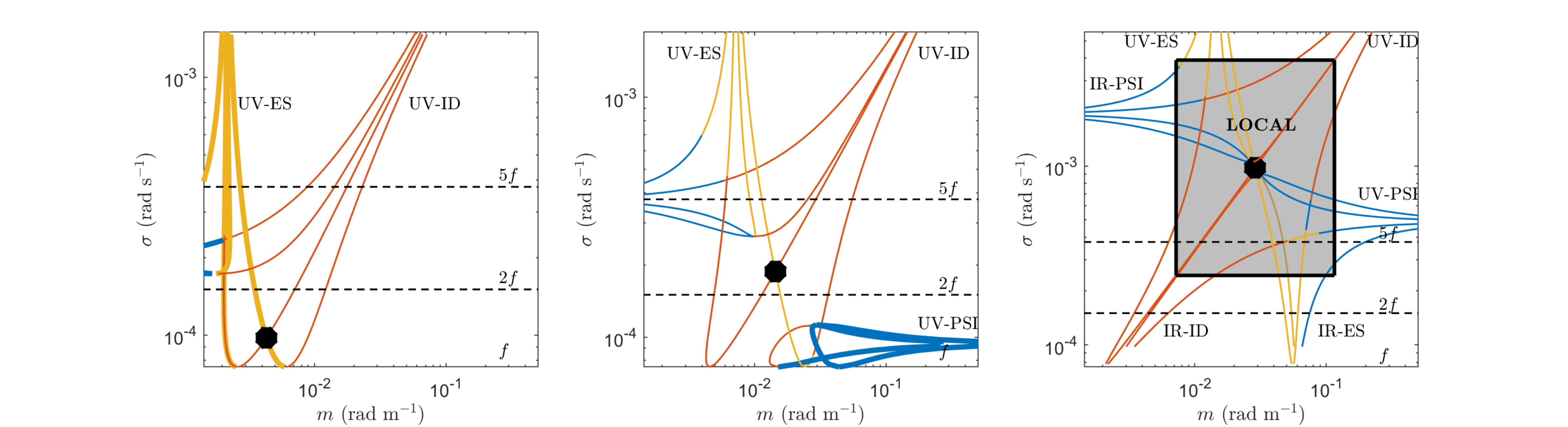}
\caption[]{The resonant manifold (\ref{ResonantManifold}) for a wave ($\bf{p},\sigma$) located at the black symbol for a) $f < \sigma  <2f$, b) $2f < \sigma <5f$ and c) $\sigma > 5f$.  The presentation in the figure assumes horizontal wavenumbers in the triad are co-linear.  Our analysis identifies the relaxation timescale $\tau_r$ as a short time scale loss (negative tendency) process.  These parts of the manifold are identified through heavier line weighting in a) and b), while they are dominated by local interactions -- contained in the shaded box in panel c) -- for higher frequencies. Here, the three asymptotic branches of the resonant manifold are labeled as Induced Diffusion (ID), Elastic Scattering (ES), and Parametric Subharmonic Instability (PSI), either in the ultraviolet (UV) or infrared (IR) limits. See the Supplementary Material of \cite{dematteis2024interacting} for specific details on how the relaxation time scale is computed.     }\label{fig:ResonantManifold}
\end{figure}

The integrand contains layers of subtractive cancellation and evaluation of the integrals requires careful consideration of convergence in extreme scale separated limits.  We lean upon the characterizations of the internal wave Kinetic Equation presented in \citep{polzin2025one,dematteis2024interacting,dematteis2022origins,dematteis2021downscale,lvov2010oceanic}.

\subsubsection{vertical relaxation time scale}
The relaxation time scale in vertical coordinate is identified by \cite{M76}.  This is a Bragg scattering (or elastic scattering) process \citep{polzin2025one} in which interaction of a wave with either shear or density variations of half the vertical wave scale and much larger horizontal scale transfer wave energy and momentum into a second wave of similar frequency, similar horizontal wavenumber and similar vertical wavenumber magnitude, but opposite sign.  This reverses the sign of the group velocity and momentum flux in (\ref{eq:verticalEnergetics}) and represents a direct damping of the energy exchange.  This also happens to be the fastest nonlinear time scale.  

There is a direct analogue to closures in Open Quantum Systems.  Here, the system Hamiltonian is the Eulerian frequency $\sigma$, the bath Hamiltonian is the energy spectrum $n^{(0)}/\sigma$ and the system-bath interaction Hamiltonian appears as equation (3.12) in \cite{lvov2024generalized} with associated closures developed through a backwards in time integration along ray trajectories, see \cite{lvov2024generalized} their section 3.2.4 and \cite{polzin2025one}, their Appendix C.  

For the GM76 spectrum, $\tau_r^{-1} = 2 k_h E_o m_{\ast}/N$, in which $E_o m_{\ast}$ is the high vertical wavenumber asymptote of the shear spectral density.  Using Appendix C of \cite{polzin2025one} we generalize $\tau_r$ to 
\begin{equation}\label{VerticalRelaxation}
\tau_r =  \frac{N^2 [m_{\ast}^2 + 4m^2]^{q/2}}{\pi m (2m)^2 B E_0 (\sigma^2-f^2)^{1/2}} ~ .
\end{equation}
in which $B$ is the normalization constant of the vertical wavenumber spectrum.   

\subsubsection{horizontal relaxation time scale}

Our intuition leads us to specifying the relaxation timescale $\tau_R$ for horizontal wavenumber as a return to the regional stationary state $n^{(0)}$, for which we provide an empirical definition in Section \ref{Model}.\ref{RegionalSpectrum}.  However, there is significant subtlety in defining this process.  From the polarization relations (\ref{eq:PolarizationRelations}), flipping the sign of horizontal wavenumber has no impact on the sign of the horizontal pseudo-momentum flux and thus nonlinear interactions that accomplish this do not represent a relaxation mechanism.  Similarly, resonant interactions are dominated by co-linear horizontal wavenumbers \citep{dematteis2021downscale}, which at first blush implies small action tendency in horizontal azimuth.  However, a two-dimensional representation of the kinetic equation involves discarding the entire last line of the collisional cross section (\ref{eq:CollisionalCrossSections}).  Interpretation of results presented in \cite{bel2022resonant} suggests that this two-dimensional representation is likely to be a singular limit.  It is with this knowledge that we offer the following conjecture:  horizontal anisotropy resulting from wave refraction in background horizontal gradients is erased at the same rate as departures from the stationary state are acted upon by loss terms in the kinetic equation.   

\begin{figure}
\includegraphics[width=16.0cm]{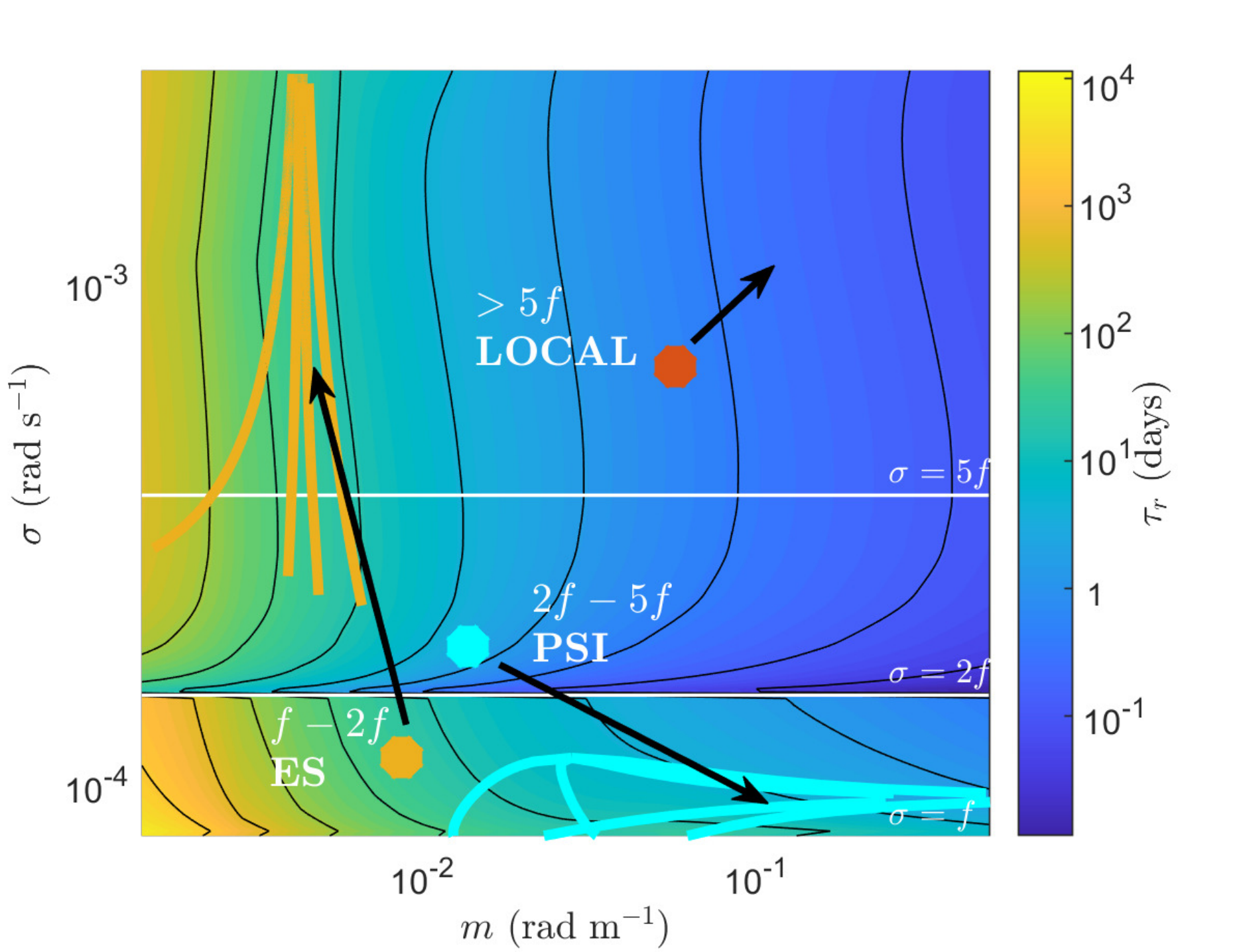}
\caption[]{Numerical evaluations of relaxation time scales $\tau_r$ following \cite{dematteis2024interacting}.  The frequency coordinate is divided into subdomains of $f-2f, 2f-5f \; {\rm and} \; 5f-N$ with caricatures of the resonant manifold from Figure \ref{fig:ResonantManifold} depicting the dominant nonlinear transfer mechanisms.  }\label{GiovanniTimeScale}
\end{figure}

Horizontal eddy-wave exchanges represent a source of wave energy throughout the vertical wavenumber - frequency domain rather than a sink.  We therefore divide the nonlinear transfers at wavenumber $\p$ into positive and negative contributions, and assign the relaxation time scale $\tau_{r}$ to those negative, or loss, contributions, figure \ref{GiovanniTimeScale}. This can be framed as a linearization of the wave kinetic equation \eqref{BroadenedKineticEquation} around the homogeneous stationary state:
\begin{equation}
    \frac{\partial}{\partial t}n^{(1)}_{\bf p}=\eta_{\bf p}(n_{\bf p}^{(0)}) - \gamma_{\bf p}(n_{\bf p}^{(0)})n^{(1)}_{\bf p}\,,
\end{equation}
with $\eta_{\bf p}>0, \gamma_{\bf p}>0$, and recognizing $\gamma_{\bf p}$ as the inverse of the relaxation time $\tau_r(\bf p)$. This is in the same spirit of the relaxation-time approximations invoked in statistical physics to derive the continuum fluid equations from the Boltzmann equation (BGK model, see e.g. \cite{saint2009hydrodynamic}), with the important conceptual difference that now the background state is the forced stationary spectrum instead of thermodynamic equilibrium. These relaxation times are calculated following the methodology presented in detail in the Supplementary Material of \cite{dematteis2024interacting}, see Eq. (1) and figure 2 therein.  We interpret the patterns as a loss at frequencies $2f \leq \sigma \leq 5f$ to $f \leq \sigma \leq 2f$ via Parametric Subharmonic Instability (PSI), loss at $f \leq \sigma \leq 2f$ to $\sigma^2 \gg f^2$ via Elastic Scattering (ES) and a local interaction at $\sigma \ge 5f$.  Connectivity in the spectral domain is indicated in figure \ref{GiovanniTimeScale} by depictions of the relevant pieces of the resonant manifold.  Here, the results presented for the horizontal relaxation time scale are based on the numerical estimates shown in figure \ref{GiovanniTimeScale}.

\section{Model Evaluation}\label{Model}
\label{Coupling}
\subsection{Navigating Algebra in M\"uller (1976)}\label{navigation}

The algebra festival in \cite{M76} relating to our $n_3^{(1)}$ is presented in $ijk$ notation, and factors involving the relaxation time scale are cloaked in the guise of an inverse functional, $D^{-1}$:  $n_3^{(1)} \rightarrow D^{-1}[k^\alpha \frac{\partial}{\partial k^m}n^{(0)}_3 \frac{\partial}{\partial x^m} \overline{u}^{\alpha}]$.  Clarity can be obtained by straightforward evaluation of the expressions for the momentum fluxes.  For the vertical coordinate:  
\begin{eqnarray}
\overline{uw} -\frac{f}{N^2}\overline{vb} & = & \int d\p \; k \; C_g^z \; n^{(1)}_3(\p; r, t) \nonumber \\
& & {\rm substituting} \; (\ref{eq:HolyGobSmack}) \nonumber \\
& = & \int d\p \; k \; C_g^z \; \frac{\tau_r}{1+(\tau_r / \tau_p)^{2}} \; \dot \p \cdot \nabla_{\p} \; n^{(0)}_3(\p) \;  \nonumber \\
&  & {\rm substituting} \; (\ref{eq:Eikonal}) \nonumber \\
& &  \dots [(-kU_x-lV_x)\partial_k +(-kU_y-lV_y)\partial_l +(-kU_z-lV_z)\partial_m ]  \; n^{(0)}_3(\p)\nonumber \\
& &  {\rm and \; using \; the \; chain \; rule \; } \nonumber \\
& & \partial_k n^{(0)}_3 =  \frac{-k}{k_h} \partial_{k_h} n^{(0)}_3  {\rm \; , \; etc.} \nonumber 
\label{VerticalMomentumManipulations}
\end{eqnarray}
This will produce a series of moments involving powers of the components of the horizontal wavevector $(k,l)$.  After changing variables of integration from $(k,l,m)$ to $(k_h,m,\phi)$, it becomes obvious that only moments in even powers of both $k$ and $l$ survive.  

Thus, by repeating the Fickian mantra that `flux equals minus a constant times a gradient', one can systematically define the `constants' relating cospectra $C(\p)$ and the background gradients.  As detailed in the Appendix of \cite{M76}:  
\begin{eqnarray}
\int C_{uu}(\p) d\p & = & -\frac{1}{2} \; \nu_h \; 2 U_x  \nonumber \\
\int C_{uv}(\p) d\p & = & -\frac{1}{2} \; \nu_h \; ( V_x + U_y) \nonumber \\
\int C_{vv}(\p) d\p & = & -\frac{1}{2} \; \nu_h \; 2 V_y  \nonumber \\
\int C_{uw}(\p) d\p & = & -\frac{1}{2} \; \nu_v \; U_z \nonumber \\
\int C_{vw}(\p) d\p & = & -\frac{1}{2} \; \nu_v \; V_z \nonumber \\
\int C_{ub}(\p) d\p & = & -\frac{1}{2} \; K_h \; B_x \nonumber \\
\int C_{vb}(\p) d\p & = & -\frac{1}{2} \; K_h \; B_y 
\label{enstrophy_as_visc}
\end{eqnarray}
The Fickian constants are given by:  
\begin{equation}
\nu_h = -\frac{1}{8} \int d\p ~ \frac{\omega^2-f^2}{\omega^2}  ~ 
\frac{\omega  ~  k_h ~\tau_R}{1+(\tau_{R}/\tau_{p})^2} \frac{\partial n_3^{(0)}(k_h,m)}{\partial k_h}~, {\rm with} \; \tau_p^{-1} = C_g^h/L  
\label{visc}
\end{equation}
and
\begin{equation}
\nu_v + \frac{f^2}{N^2}K_h = \frac{1}{2} \int d\p ~ \frac{\omega^2-f^2}{\omega^2}  ~ 
\frac{\omega  ~  k_h^2}{m} \frac{\tau_R}{1+(\tau_{R}/\tau_{p})^2} \frac{\partial n_3^{(0)}(k_h,m)}{\partial m}~, {\rm with}\; \tau_p^{-1} = C_g^z/H.
\label{visc-vert}
\end{equation}
using single scales $L \gg k_h^{-1}$ and $H \gg m^{-1}$, consistent with the extreme scale separated paradigm.  

The Fickian `constants' (\ref{visc}) and (\ref{visc-vert}) capture the lowest order coupling between nominally `slow' eddies and `fast' internal waves.  While the algebra is time consuming, there is a straight forward path for considering higher order effects via introducing background spatial gradients in the dispersion relation that propagate through the eikonal relations (\ref{eq:Eikonal}) and departures from strict geostrophy such as `gradient wind' and frontogenetic conditions, Section \ref{Summary}.\ref{Optics}.\ref{WKB}.  See the Appendix of \cite{Polzin96a} for cautionary notes.  

We will present evaluations of (\ref{visc}) and (\ref{visc-vert}) in Section \ref{Coupling}.\ref{Evaluations} after defining the regional spectrum $n_3^{0}$ (Section \ref{Coupling}.\ref{RegionalSpectrum}), propagation time scales $\tau_p$ (Section \ref{Coupling}.\ref{PropagationTimeScales}) and relaxation time scales $\tau_r$ (Section \ref{Prognostic}.\ref{Damping}).

\subsection{The Regional Spectrum}\label{RegionalSpectrum}

Here we introduce the regional spectrum for the Sargasso Sea, where the Local Dynamics Experiment was located.  This spectrum defines the background $n^{(0)}(\p)$ in our calculations.  We do so with the intent of underscoring the hypothesis of \cite{polzin2011toward} that variability in parametric spectral fits to internal wave frequency and vertical wavenumber reflect regional variability in the sources and dominant nonlinear mechanisms.  The Sargasso Sea is south of the Gulf Stream and thus south of the east coast storm track that couples strongly to near-inertial motions.  Coherent tidal propagation is moderate.  See figures 45 and 47 of \cite{polzin2011toward}.  The Sargasso Sea paradigm articulated in \cite{polzin2010mesoscale} is that transfers from the mesoscale to the internal wavefield, diagnosed by direct estimates of kinetic energy transfers, can balance the observed turbulent dissipation rates inferred via finescale parameterization methods.  This Sargasso Sea spectrum can thus be regarded as an end member in (tidal, atmosphere, eddy) forcing parameter space that represents the mesoscale eddy - internal wave coupling process.  The point of interpretation is that coupling with the mesoscale acts as an amplifier of tertiary inputs by smaller sources at lower vertical wavenumber than are captured by the parametric spectral fits and that these Sargasso Sea spectral parameters are characteristic of eddy-wave coupling.  

We use 
\begin{equation}\label{ParametricSpectralRepresentation}
E_2^{(0)}(\sigma,m) = \frac{B E_0 }{(m_{\ast}^2+m^2)^{q/2}} \frac{C}{\sigma^{p-2r} (\sigma^2-f^2)^{r}}
\end{equation}
to characterize the internal wave spectrum.  The factor $C$ normalizes the frequency spectrum to 1.0.  The factor $B$ represents the normalization constant for the vertical wavenumber spectrum and the total energy is $E_0$.  

\begin{figure}\hspace{-2.0cm}
\includegraphics[width=18.0cm]{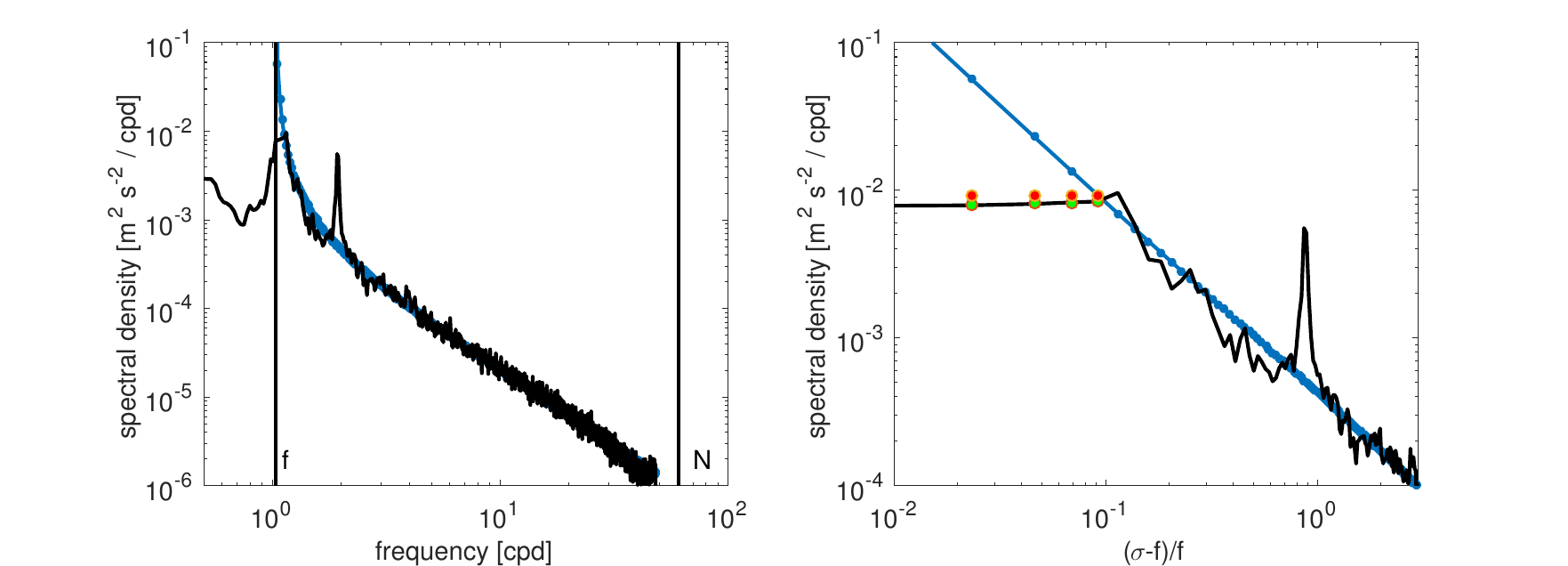}
\caption[]{600 m LDE frequency spectra of horizontal velocity with fits.  Black traces and green circles represent the observations, blue traces the parametric fits according to formula \eqref{ParametricSpectralRepresentation}.  The red circles represent the regularization of the parametric curve. Notice that the second panel shows the same curves but with a rescaling of the horizontal axis that emphasizes the near-inertial peak.}\label{FrequencySpectra}
\end{figure}

Frequency spectra from current meters at 616 and 839 m water depth on the center mooring of the LDE array, Figure \ref{FrequencySpectra}, are used to document the total internal wave energy ($E_{0} = 0.00192$ m$^2$ s$^{-2}$, and 0.00172 m$^2$ s$^{-2}$, respectively).  Stratification profiles from CTD casts taken during the experiment provide the environmental metrics $N^2 = 1.94\times10^{-5}$s$^{-2}$ at 600 and $N^2 = 2.29\times10^{-5}$s$^{-2}$ at 825 m, and $\int_{{\rm bottom}}^{{\rm surface}} N(z) dz / 0.00524 \; $s$^{-1}$ = 1830 m rather than the nominal 1300 m scale height of the GM model.  Curve fitting the frequency spectra leads us to $p=1.70$ and $r=1.33$.  We follow \cite{dematteis2024interacting} in regularizing the frequency integral for nominally non-integrable singularities by introducing a cutoff frequency $\sigma_{{\rm cut}}$ with a bandwidth of $0.09f$.  
The spectral parameters are characteristic of the region and are repeatable through a plethora of observational campaigns:  the Mid-Ocean Dynamics Experiment, the Internal Wave Experiment, PolyMODE I, II, and III, and the Frontal Air-Sea Interaction Experiment, as discussed in \cite{polzin2011toward}.

Data that inform spectral parameters in vertical wavenumber are not as common and often come with the caveat that they are biased due to ship-board sampling strategies that target coherent features rather than attempting to document the background internal wavefield.  This is especially true of Sanford's MODE \citep{LS75, polzin2008mesoscale} and LDE \citep{ES86b} data sets.  However, the gamut of data sets (including the Fine-and-Microstructure Experiment \citep{gargett1981composite}, IWEX \citep{M78} and and FASINEX \citep{Polzin96a}) leads to a regional picture of relatively red vertical wavenumber spectra with wavenumber cut-off $m_c$ slightly smaller than 0.1 cpm.  From this regional perspective we argue for $q=2.25$ and for the integral of the shear spectral density $\int_0^{m_c} 2 E_k(m) dm = 2\pi/10 \; N^2$ at $m_c = 0.080$ cpm.  We then numerically iterate to find the value of $m_{\ast}$ that provides the total wave energy from the LDE current meter.  This process leads us to $m_{\ast} = 0.0194$ m$^{-1}$ at 600 m, and $m_{\ast} = 0.0245$ m$^{-1}$ at 825 m.  After buoyancy scaling to $N_0 = 3$ cph, these are equivalent to approximately mode 14.5 (13.5 and 15.6, respectively) in our 1830 m deep ocean.  Our curve fits are summarized in Table \ref{ParametricSpectralParameters}.  

\begin{table}[h!]
\normalsize
\centering
\caption{Results from Parametric Spectral Fitting}\label{ParametricSpectralParameters}
\begin{tabularx}{0.8\textwidth}{r c c r}
nominal depth [m] & \;\;\;\;\;\;\;\; 600 \;\;\;\;\;\; & \;\;\;\;\;\; 825 \;\;\;\;\;\;\;\; & GM \\  
$N^2$ \; $[\rm{s}^{-2}]$ & $1.94\times10^{-5}$ & $2.29\times 10^{-5}$ & $2.75\times10^{-5}$ \\
$E_{0}$ \; $[{\rm m}^2$ \; ${\rm s}^{-2}$]  & $19.2\times10^{-4}$ \; & $17.2\times10^{-4}$ & $30\times10^{-4}$ \\
$q$ & 2.25 & 2.25 & 2.0 \\
$p$ & 1.70 & 1.70 & 2.0 \\
$r$ & 1.33 & 1.33 & 0.50 \\
$m_{\ast}$ \;\;  $[{\rm m}^{-1}]$ & 0.0194 &  0.0245 & 0.0069 \\
$E_k/E_p$ & 3.84 & 3.53 & 3.0 \\ 
\end{tabularx}
\end{table}


There are several points to emphasize:  
\begin{itemize}
\item The bandwidth $j_{\ast}$ is not 1, nor 4.  It is the equivalent of mode 15.  We are {\em not} talking about waves with horizontal group velocities of $O(2)$ m s$^{-1}$, which are the internal wave equivalent of surface wave swell.  The parametric fitting describes internal waves with horizontal group velocities of $O(0.1)$ m s$^{-1}$ that will be strongly modulated by interaction with the LDE mesoscale velocities of $O(0.1)$ m s$^{-1}$.  The parametric spectral model describes the internal wave equivalent of wind waves.  
\item The vertical wavenumber / frequency power laws are {\em not} in the combination $p=q$.  Thus, unlike the GM76 model, the 3-d action spectrum, $n_3^{(0)}(m,k_h) \propto m^{p-q} k_h^{-p-2}$, is {\em not} independent of vertical wavenumber.  This leads to efficient vertical coupling of high frequency internal waves with thermal wind shear.  
\end{itemize}

\subsection{Propagation Time Scales}\label{PropagationTimeScales}

\underline{vertical scale $H$}\\
\cite{Munk66} buoyancy scales to $N=3$ cph to determine  $H=1300/\pi$ m just west of San Diego.  Using the same approach in the Sargasso Sea provides something closer to $H=1830 $m$ / \pi=580$ m.  See Figure \ref{model_vert_H} of Appendix B for model sensitivity to this choice. 

\underline{horizontal scale $L$}\\
\cite{Owens85}, their Figure 9, estimates a zero crossing of the transverse velocity correlation function of 100 km from the LDE current meter data. The longitudinal velocity correlation function falls off more slowly and thus the longitudinal length scale is not resolved.  We will use $L = 100$ km.  
See Figure \ref{model_horz_L} of Appendix B for model sensitivity to this choice.  

\underline{Timescale $\Omega$}\\
The primary concern as regards the timescale $\Omega$ is the propensity to set up a resonance in which the group velocity of the internal wave equals the phase speed of the background, $\Omega = {\bf K} \cdot {\bf C}_g$.  The importance of a resonance is that it provides a very long timescale over which energy exchanges can occur.  While observed mesoscale phase speeds are small, on the order of 2 cm s$^{-1}$ \citep{McW76}, this resonance has been identified \citep{polzin2008mesoscale} in the context of Sanford's data from the MODE experiment \citep{LS75}, slightly to the south.  Time dependence is greater at the LDE site in conjunction with what were identified as Topographic Rossby Waves \citep{PR82} with phase speed of 0.06 cm s$^{-1}$.  The resonance is, without doubt, of some relevance to sustaining the near-inertial wave field.  However, the presence of a resonance is not the primary concern, {\it per se}, in this paper as the issue of westward phase propagation adds a layer of complexity.  Our focus is on identifying and documenting the damping process.    


\begin{figure}
\includegraphics[width=0.75\textwidth]{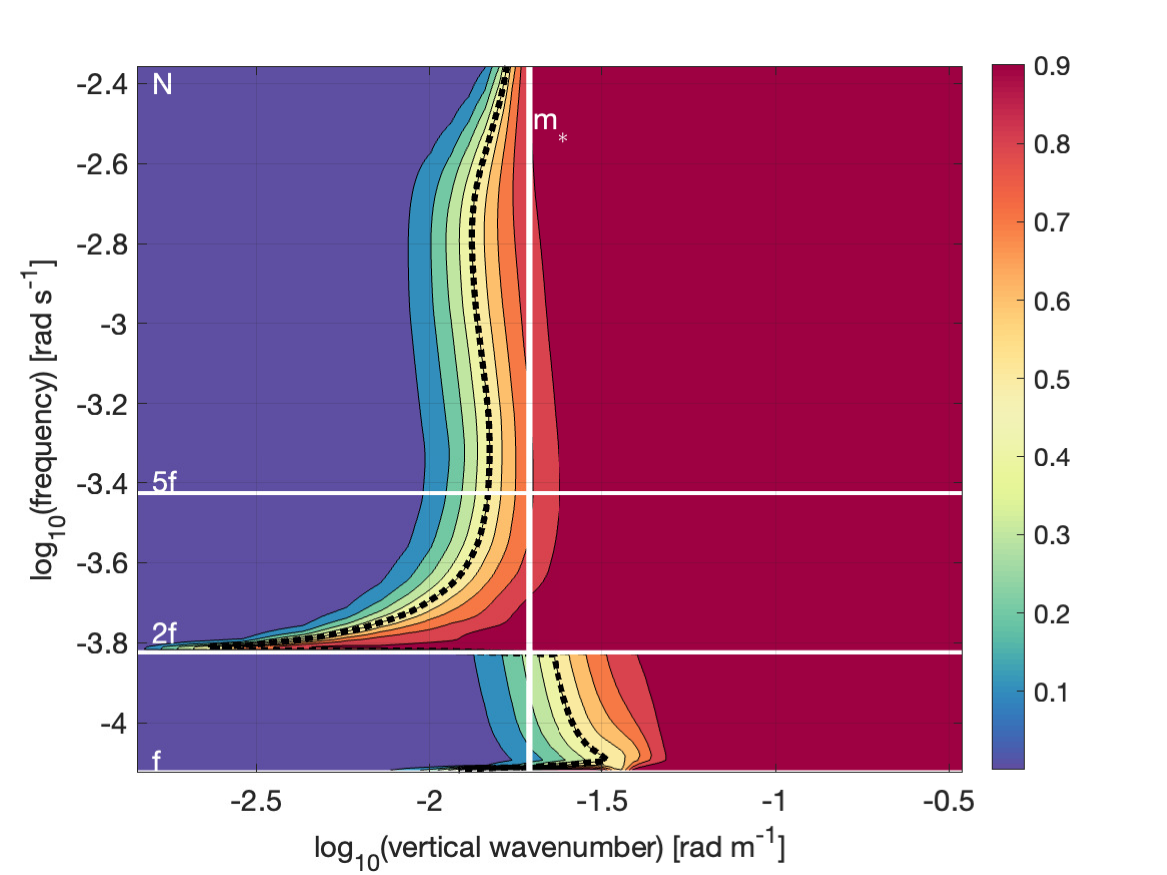}
\caption[]{ The transfer function $1/(1+(\tau_R/\tau_p)^2)$ for the horizontal process.  The dashed black line indicates the half-power point where $\tau_R=\tau_p$.  The structure reflects the nonlinear relaxation timescales apparent in Figure \ref{GiovanniTimeScale} and the vertical wavenumber bandwidth $m_{\ast}$.  }\label{TimeScalesTFN}
\end{figure}

\subsection{Data Comparison}\label{Evaluations}
\subsubsection{The Horizontal Dimension}

In the hydrostatic approximation, the horizontal viscosity (\ref{visc}) becomes:
\begin{equation}
\nu_h = -\frac{1}{8} \int d\p ~ \frac{\omega^2-f^2}{\omega^2}  ~ 
\frac{\omega  ~  k_h ~\tau_R}{1+(\tau_{R}/\tau_{p})^2} \frac{\partial n_3^{(0)}}{\partial k_h}~,
\end{equation}
in which $k_h$ represents the magnitude of the horizontal wavevector components, $k_h = (k^2 +l^2)^{1/2}$.  The relaxation rate $\tau_r$ appears in figure \ref{GiovanniTimeScale}, the propagation timescale $\tau_p^{-1}= C_g^h/L$ and the transfer function appears in Figure \ref{TimeScalesTFN}.  

The relation between the 2-D energy spectrum, Section \ref{Model}.\ref{RegionalSpectrum}, and a 3-D isotropic action spectrum dictates:
\begin{equation}
n_3^{(0)}(m,\omega) = E_2^{(0)}(m,\omega) / k_h \omega ~.
\end{equation}
The parametric representation (\ref{ParametricSpectralRepresentation}) provides: 
\begin{equation}
\frac{\partial n^{(0)}_3}{\partial k_h} = \frac{N^3 B C m^{-3}}{(m_{\ast}^2+m^2)^{q/2}} \frac{1}{(\sigma^2-f^2)^{r+1/2}} (\frac{1}{\sigma^2})^{(2-r+p/2)} [-\sigma^2(2+p) +f^2(2-2r+p)] \; .
\label{BarnFarts}
\end{equation}
Regularization (see section \ref{Coupling}.\ref{RegionalSpectrum}) for frequencies $\sigma \leq \sigma_{{\rm cut}}$ is handled by 
\begin{equation}
\frac{\partial n^{(0)}_3}{\partial k_h} = E_2^{(0)}(m,\sigma_{{\rm cut}}) \frac{\partial}{\partial k_h} (\frac{1}{k_h \sigma} \frac{\partial \sigma}{\partial k_h})
\nonumber
\end{equation}
We evaluate (\ref{BarnFarts}) by integrating over horizontal azimuth, in which 
\begin{equation}
\int d\p \rightarrow \frac{1}{2\pi} \int_0^{2\pi} d \phi \; k_h \; dk_h dm \nonumber
\end{equation}
and then transform from $(k_h, m)$ space to $(\sigma, m)$ space:
\begin{equation}
\nu_h = -\frac{1}{8} \int d\sigma dm \frac{m^3}{N^3} (\sigma^2 - f^2)^{3/2} \frac{\tau_r}{1+(\frac{\tau_r}{\tau_p})^2} \frac{\partial n^{(0)}_3}{\partial k_h} \; . 
\label{nuHofmw}
\end{equation}
Numerical evaluation returns
\begin{equation}
\nu_h \cong 50 ~{\rm m}^2 {\rm s}^{-1} ~
\end{equation}
for Sargasso Sea parameters.

\begin{figure}
\includegraphics[width=35pc]{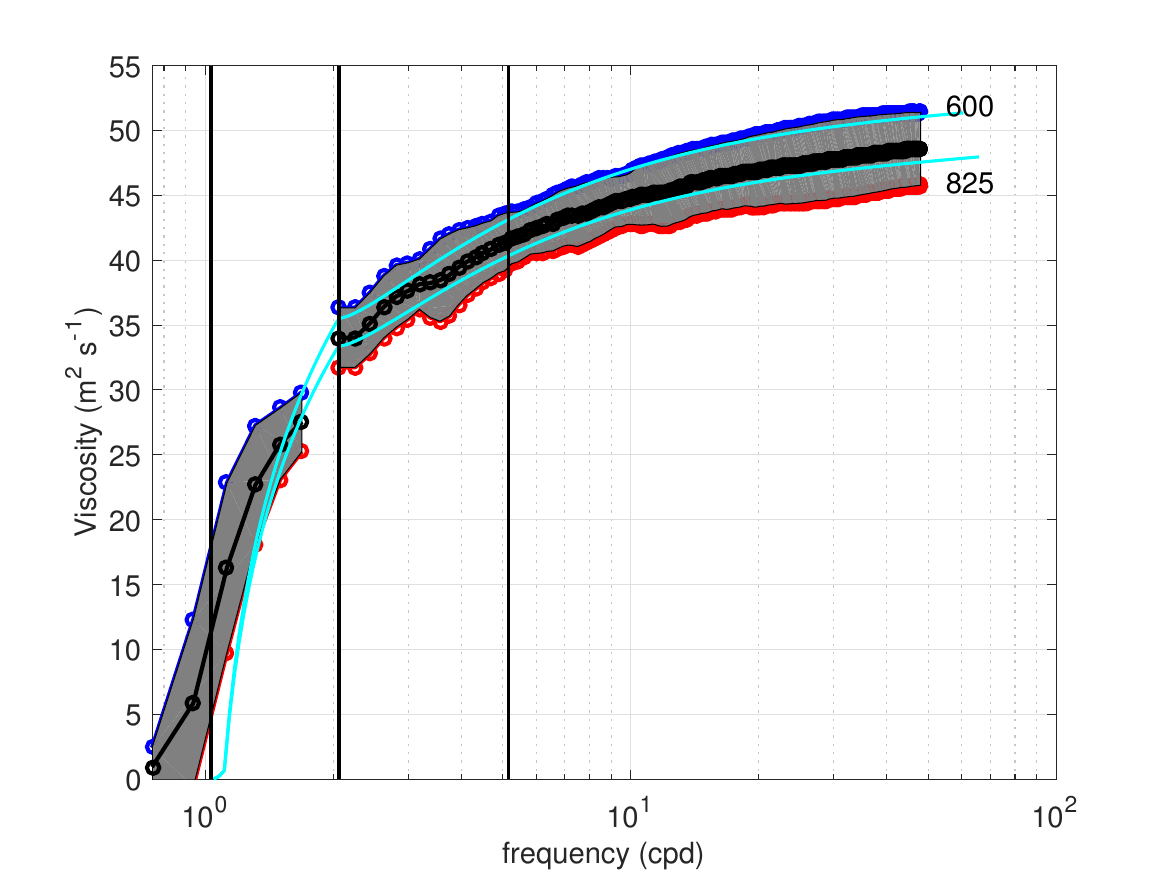}
\caption[]{Cumulative integrals of the spectral functions $-sgn(S_n)[P_{u^{\prime\prime}u^{\prime\prime}}-P_{v^{\prime\prime}v^{\prime\prime}}]$ (blue dots) and $-2sgn(S_s) C_{u^{\prime\prime}v^{\prime\prime}}$ (red dots), divided by rate of strain $\overline{\overline{ \mid S_n \mid }}$ and $\overline{\overline{ \mid S_s \mid }}$, to provide estimates of the horizontal viscosity $\nu_h$.  Blue and red dots are connected by grey shading and the mean is represented as black dots.  Over plotted as cyan lines are Sargasso Sea based estimates for the corresponding viscosity estimate using (\ref{nuHofmw}) with parameters estimated at 600 and 825 m.  These estimates employ an eddy scale of $L =$ 100 km.  The vertical black lines delineate frequencies of $f$, $2f$ and $5f$.  }\label{diff_h}
\end{figure}

We integrate over vertical wavenumber and compare the cumulative frequency integral of (\ref{nuHofmw}) with the observed coherence functions normalized by the respective rate of strain, Fig. \ref{diff_h}.  In this instance semidiurnal frequencies make a significant contribution to the shear component of the stress -- rate of strain relation \citep{polzin2010mesoscale}.  These contributions are assumed to be associated with low-vertical modes that are not described by the parametric fits (\ref{ParametricSpectralRepresentation}). Thus, the comparison is made by excluding semi-diurnal frequencies from the cumulative integrals of the observed coherence functions.  The observations are adjusted through this gap so as to aid the eye.  In so doing, the `observed' frequency integrated estimates of horizontal viscosity of 50 m$^2$ s$^{-1}$ agree with theoretical estimates using a length scale of $L = 100$ km.  We begin integration of the observed coherence functions at $\sigma = 0.7 f$ rather than at $f$.  This includes some covariance at frequencies $\sigma < f$ associated with Doppler smearing that is not included in the numerical assessments of (\ref{nuHofmw}) which utilize the intrinsic frequency $\omega$.  This Doppler smearing along the frequency axis exceeds that associated with modulation of the near-inertial waveguide by relative vorticity by an order of magnitude.  These two differences in methodology result in figure \ref{diff_h} having a slightly different appearance in comparison with figure 3 of \cite{polzin2010mesoscale}

Uncertainty estimates are presented in \cite{polzin2010mesoscale} for the stress--strain regressions.  They are far larger than the $O(10$--$20 \%)$ mismatch between data and theory and render explanations of slight apparent disparities between data and theory to be speculative, at best. See Appendix B for uncertainty in the underlying coherence estimates, sensitivity to the Regional parametric spectral representation and sensitivity to mesoscale length scale.    

\subsubsection{The Vertical Dimension}

In the hydrostatic approximation, the effective vertical viscosity (\ref{visc-vert}) becomes:
\begin{equation}
\nu_v + \frac{f^2}{N^2}K_h = \frac{1}{2} \int d\p ~ \frac{\omega^2-f^2}{\omega^2}  ~ 
\frac{\omega  ~  k_h^2}{m} \frac{\tau_R}{1+(\tau_{R}/\tau_{p})^2} \frac{\partial n_3^{(0)}(k_h,m)}{\partial m}~, 
\end{equation}
with relaxation rate $\tau_r$ given by (\ref{VerticalRelaxation}) and propagation timescale $\tau_p^{-1} = C_g^z/H$.  

For the parametric spectral representation (\ref{ParametricSpectralRepresentation}), 
\begin{equation}
\frac{\partial n^{(0)}}{\partial m} = N^2 B C (\frac{1}{m^2(\sigma^2-f^2)})^r \frac{m^p}{(m_{\ast}^2 + m^2)^{q/2}} (\frac{1}{m^2\sigma^2})^{1-r-p/2} \big[\frac{-qm}{(m_{\ast}^2+m^2)} + \frac{p}{m} - 2\frac{f^2}{\sigma^2}\frac{(1-r+p/2)}{m}\big]
\end{equation}

\begin{figure}
\includegraphics[width=35pc]{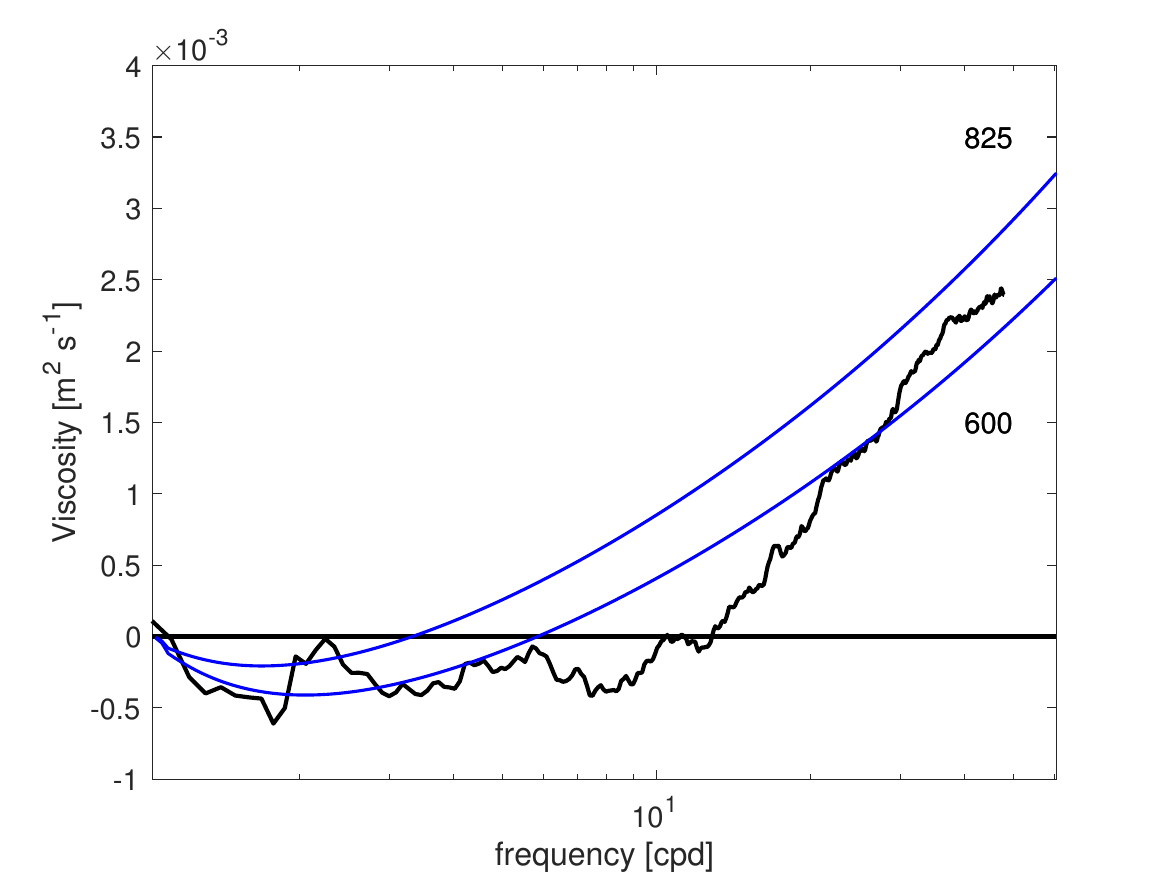}
\caption[]{Cumulative integrals of the spectral function $-sgn(\overline{u}_z)[C_{u^{\prime\prime}w^{\prime\prime}} - fC_{v^{\prime\prime}b^{\prime\prime}}/N^2]/ (\overline{u}_z^2)^{1/2}  - sgn(\overline{v}_z)[C_{v^{\prime\prime}w^{\prime\prime}} + fC_{u^{\prime\prime}b^{\prime\prime}}/N^2]/ (\overline{v}_z^2 )^{1/2}$, to provide estimates of the vertical viscosity $(\nu_v + \frac{f^2}{N^2} K_h)$.  Over plotted as thick lines are prognostic estimates (\ref{nuVofmw}) derived from current meters at 600 and 825 m water depth.  } \label{diff_v} 
\end{figure}

Regularization (see section \ref{Coupling}.\ref{RegionalSpectrum}) for frequencies $\sigma \leq \sigma_{{\rm cut}}$ is handled by 
\begin{equation}
\frac{\partial n^{(0)}_3}{\partial m} = \frac{\partial}{\partial m} (\frac{1}{k_h \sigma} \frac{\partial \sigma}{\partial k_h} E_2^{(0)}(m,\sigma_{{\rm cut}}) )
\nonumber
\end{equation}
We similarly integrate over horizontal azimuth and transform to $(\sigma, m)$ space to obtain:   
\begin{equation}
\nu_v + \frac{f^2}{N^2}K_h = \frac{1}{2} \int d\sigma dm  ~ \frac{\big[ \omega^2-f^2 \big]^2 m^3}{N^4} \frac{\tau_R}{1+(\tau_{R}/\tau_{p})^2} \frac{\partial n^{(0)}_3}{\partial m}~.
\label{nuVofmw}
\end{equation}
Numerical evaluation for the Sargasso Sea spectral parameters provides $\nu_v + \frac{f^2}{N^2}K_h = 2.5\times 10^{-3}$ m$^2$ s$^{-1}$.  After integration of (\ref{nuVofmw}) over vertical wavenumber, direct comparisons with the cumulative contributions to the frequency integral (figure 6 of \cite{polzin2010mesoscale}) are presented in figure \ref{diff_v}.  

The observed contributions (figure \ref{diff_v}) are negative at near-inertial frequencies and trend positive at high frequency, indicating near-inertial wave energy loss to the mean flow consistent with critical layer behavior and energy gain within the continuum, $\sigma^2 >> f^2$.  The observed contributions at high frequency are consistently positive at $\sigma > 10$ cpd.  

\begin{figure}
\includegraphics[width=35pc]{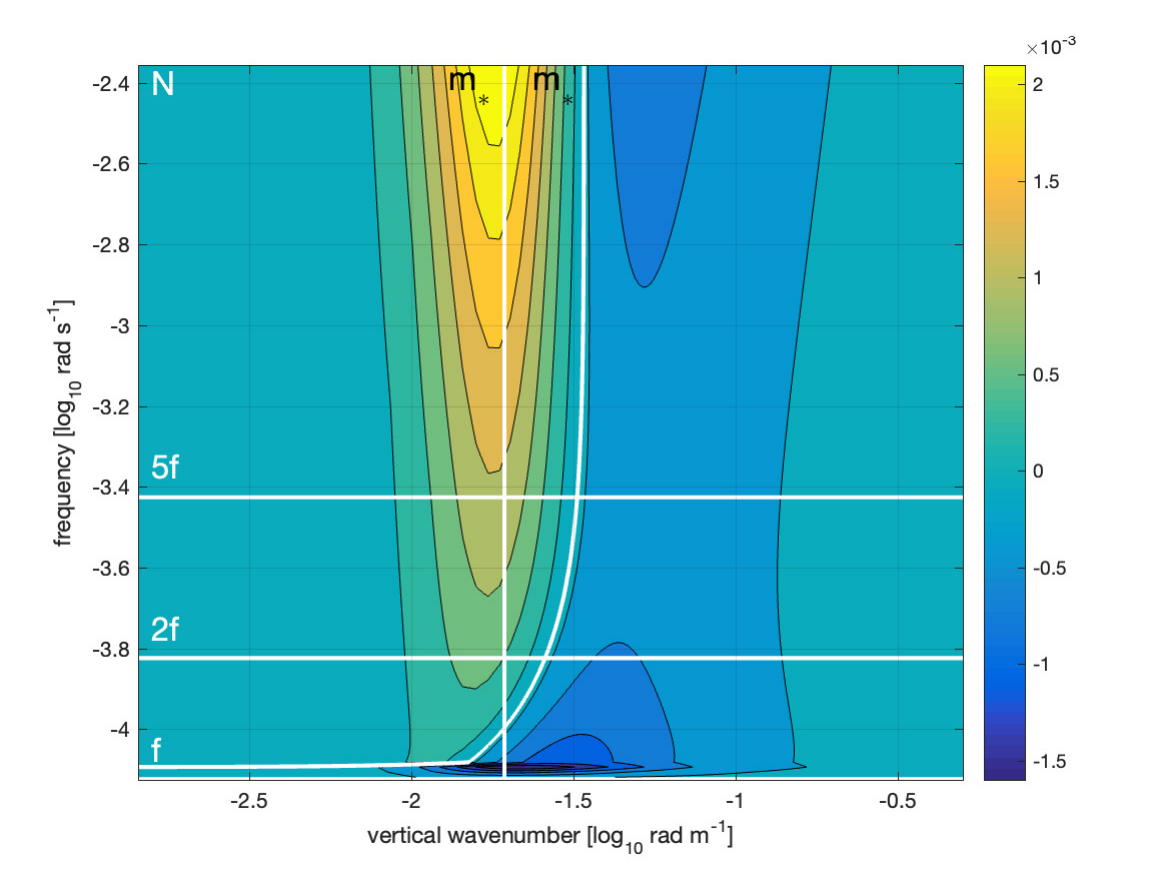}
\caption[]{The integrand of (\ref{nuVofmw}) multiplied by vertical wavenumber $m$ and frequency $\sigma$.  The sign transition (white contour) is associated with change in sign of the gradient of the background spectrum, $\partial_m n_3^{(0)}({\p})$, which in turn is determined by  the bandwidth $m_{\ast}$ in the non-rotating ($\sigma^2 \gg f^2$) limit.  The sign after integration over vertical wavenumber is sensitive to the interplay between the bandwidth $m_{\ast}$ and the non-dimensional ratio $\tau_r/\tau_p$, such that the zero crossing of the cumulative viscosity in figure (\ref{diff_v}) requires accurate specification of both.  If, for example, one used the GM76 model, the comparison in figure (\ref{diff_v}) is unrealistic, Appendix B.  See also figures 42 and 43 of \cite{polzin2011toward}. }\label{integrand_v}
\end{figure}

The frequency domain summary in figure (\ref{diff_v}) is a result of subtractive cancellation in the integrand of (\ref{nuVofmw}), with positive coherence indicating wave extraction of energy from the mean flow at large ($m < m_{\ast}$) vertical scales and energy loss to the mean flow at smaller ($m > m_{\ast}$) vertical scales, figure \ref{integrand_v}.  This change in sign results from the change in sign of the vertical wavenumber gradient of the three dimensional background action spectrum, $n_3^{(0)}$, about $m=m_{\ast}$.  We interpret this sensitivity to $m_{\ast}$ as symptomatic of the fundamental role of wave-mean interactions in setting the character of the internal wave spectrum in the Sargasso Sea.  It provides a conceptual underpinning of why the observed lowest-order description of oceanic internal wave spectra (\ref{ParametricSpectralRepresentation}) can be consistent with separability in vertical wavenumber and frequency.  See Appendix B for uncertainty in the underlying coherence estimates, sensitivity to the Regional parametric spectral representation and sensitivity to mesoscale length scale.   

In summary, the ability of the prognostic estimates to capture both the qualitative and quantitative character of the observations is remarkable given the following:  (1) The prognostic estimates are based upon a 2-d vertical wavenumber intrinsic frequency spectrum inferred from two 1-d spectra assuming separability in vertical wavenumber and Eulerian frequency.  (2) Vertical velocity was inferred from a 1-d (vertical) temperature balance.  Contamination by both vertical and horizontal Doppler shifting is unconstrained.  Both issues could be addressed from a single mooring by higher resolution temperature measurements and higher precision 3-D acoustic travel time current meters at higher vertical resolution.  

\subsection{Evaluation Summary}
After assembling the considerable algebra required to evaluate (\ref{eq:HolyGobSmack}), we have found that that the predicted energy transfer is of an appropriate magnitude and frequency distribution to match empirical estimates from the LDE field program.  In the next section we turn to placing this framework into the regional mesoscale eddy energy and potential enstrophy budgets.

\section{Budgets}\label{budgets}
In this section we frame the implications of the wave-eddy viscosity coefficients identified in the previous section for mesoscale dynamics.  This is enabled by invoking a Reynolds decomposition 
into a quasi-geostrophic `mean' (in upper case) and small amplitude internal wave (in lower case) perturbations on the basis of a time scale separation:  $\psi \rightarrow \Psi
+ \psi$ with an explicit averaging timescale $\overline{\psi} \rightarrow \tau^{-1} \int_{0}^{\tau} \psi dt$ much longer than the internal wave time scale but smaller than the eddy time scale.  We will also require a further time (or ensemble) average which , in practice, is the record length mean.  This average is represented as $\langle \dots \rangle $.  

\subsection{Internal Wave Energy}\label{IWenergy}
Direct evaluation \citep{polzin2010mesoscale} of the horizontal stress - horizontal rate of strain terms in (\ref{Energy2}) returns an energy transfer rate of $3\times10^{-10}$ W kg$^{-1}$.  Similar evaluation of vertical stress - vertical shear returns an energy transfer rate of $1\times10^{-10}$ W kg$^{-1}$.  This energy transfer rate is an $O(10)\%$ contribution to the mesoscale eddy energy budget presented in \cite{B82}, in which production through baroclinic instability and the rate of work against the mean are the leading order terms, at some $3.3\pm1.7 \times10^{-9}$ and $1.5\pm0.9 \times10^{-9}$ W/kg, respectively.  The energy transfer {\it into} the internal wavefield, however, is in balance with internal wave dissipation inferred from finescale metrics, \citep{polzin2010mesoscale}.  

\subsection{Potential Vorticity}\label{Potential_Vorticity}
The linear quasigeostrophic potential vorticity equation is [\cite{M76}]:
\begin{eqnarray}
(\partial_t + U\partial_x + V\partial_y)
(\partial_x^2 \Phi + \partial_y^2 \Phi + \partial_z[\frac{f^2}{N^2}
\partial_z \Phi] + \beta y) = \nonumber\\
\partial_x[\partial_x \overline{vu} + \partial_y \overline{vv} + \partial_z(\overline{vw} + \frac{f}{N^2}\overline{bu})] \nonumber\\
-\partial_y[\partial_x \overline{uu} + \partial_y \overline{uv} + \partial_z(\overline{uw} - \frac{f}{N^2}\overline{bv})] + {\mathcal H} 
\label{qg_pv}
\end{eqnarray}
in which ${\mathcal H} $ represents modification of the mean buoyancy profile through diabatic processes, $\Phi$ is the geostrophic streamfunction $(\Phi= \Pi/f$ with Coriolis parameter $f$) and
pressure $\Pi$ is defined in the absence of the internal
wavefield.  For parcels that do not make excursions into the mixed layer, 
${\mathcal H}  = \partial_{z} \frac{f}{N^2} \partial_z (K_{\rho} \langle B_z \rangle )$.  In the background wavefield, $K_{\rho} \cong 5\times10^{-6}$ m$^{2}$ s$^{-1}$.  Scaling terms on the right-hand-side of (\ref{qg_pv}) using this diapycnal diffusivity and the viscosity operators identified in the previous section along with estimates of vertical shear, deformation rate of strain and the first baroclinic deformation radius for a lateral scale, one finds that terms involving internal wave momentum are one to two orders of magnitude larger than that involving the dipaycnal process.  Diabatic effects are thus excluded from consideration below.  

The right-hand-side of (\ref{qg_pv}) describes the $O(\ell/{\mathcal L})$ potential vorticity signatures on the wave packet scale associated with $n_3^{(2)}$, (\ref{eq:Decomposition}).  We will assume that such contributions can be benched marked by the $n_3^{(1)}$ viscosity operators below.

\subsection{Potential Enstrophy}\label{Potential_Enstrophy}
In this subsection we tersely present a treatment of potential enstrophy transfers that parallels our treatment of energy transfers.  
\subsubsection{The Enstrophy Equation }
Including a schematic representation of eddy--eddy nonlinear transfers as ``$\mathcal{N}\mathcal{L}$", the eddy potential enstrophy budget is obtained by first decomposing the low frequency field into mean $\langle Q \rangle$ and mesoscale eddy components $Q^{\prime}$.  Multiplying (\ref{qg_pv}) by the quasigeostrophic perturbation potential vorticity and averaging returns:
 \begin{eqnarray}
 \frac{1}{2}  \frac{ D}{Dt} ~ \langle Q^{\prime 2} \rangle + 
 \langle{\mathcal N}{\mathcal L}\rangle &+& \langle {\bf U^{\prime} } Q^{\prime} \rangle \cdot \nabla_h \langle Q \rangle = \nonumber \\
 && - \langle Q^{\prime} ~ 
[ -\partial_x \nabla \cdot \int d\p ~ n({\bf p},{\bf x},t) ~ l ~ {\bf C_{g}}
+ \partial_y \nabla \cdot \int d\p ~ n({\bf p},{\bf x},t) ~ k ~ {\bf C_{g}} ] \rangle \nonumber \\
&& \label{enstrophy}
 \end{eqnarray}
in which $\frac{ D }{Dt}$ represents the material derivative following the $\langle {\rm geostrophic} \;  {\rm flow} \rangle $ and $\nabla_h$ is the 2-D horizontal gradient operator.  

\subsubsection{Viscous Closures from Wave Action (\ref{enstrophy_as_visc})}
{\em If} closure of mesoscale eddy - internal wave coupling through flux gradient relations can be justified, in which 
$-2\overline{ uv } = \nu_h (V_x + U_y)$,  
$-\overline{uw} = \nu_v U_z$, 
$-\overline{uu} = \nu_h U_x$, 
$-\overline{vv} = \nu_h V_y$, 
$-\overline{ub} = K_h B_x$, and 
$-\overline{vb} = K_h B_y$, 
considerable simplification results.  The right-hand-side of the enstrophy equation, using the thermal wind relation and using either the chain rule or integrating by parts, can be rewritten as:  
\begin{eqnarray}
  \frac{1}{2} \frac{ D}{dt} ~ \langle Q^{\prime 2} \rangle &+& 
 \langle {\bf U^{\prime}}Q^{\prime} \rangle \cdot \nabla_h \langle Q \rangle + \langle {\mathcal N}{\mathcal L} \rangle = \nonumber \\
&& - \frac{1}{2} \nu_h[\langle \zeta_x^{\prime 2} + \zeta_y^{\prime 2} \rangle + \frac{f_o^2}{\langle B_z \rangle } \langle \zeta_z^{\prime 2}\rangle  ] 
-  [\nu_v + \frac{f^2}{N^2}K_h] [\langle \zeta_z^{\prime 2} \rangle + \frac{1}{\langle B_z \rangle} \langle B_{xz}^{\prime 2} + B_{yz}^{\prime 2} \rangle ]~.  \nonumber \\
&&
\end{eqnarray}
Estimating potential enstrophy dissipation becomes an issue of estimating the gradient variance associated with potential enstrophy and the viscosity acting at the horizontal scales containing that gradient variance.

\label{Energy_Budget}

\subsubsection{LDE Potential Enstrophy Production Estimates }
Our first step in assessing potential enstrophy dissipation by coupling to internal wave momentum is to estimate the rate of potential enstrophy production.  \cite{BOB86} use the LDE array data to estimate the eddy thickness 
[$\eta^{\prime}=f_0(\rho^{\prime}/ \langle \rho_z \rangle )_z$] and 
relative vorticity ($\zeta^{\prime}$) fluxes:
\begin{eqnarray}
\langle {\bf U}^{\prime}\eta^{\prime} \rangle & = & (-0.79 \pm 0.53 ,\;  -1.45 \pm 0.71 ) \times10^{-7} ~{\rm m~s}^{-2}\nonumber\\
\langle {\bf U}^{\prime}\zeta^{\prime} \rangle  & = & (-1.57 \pm 1.51 ,\;  1.95 \pm 1.54 ) \times10^{-8}  ~{\rm m~s}^{-2} . \nonumber
\end{eqnarray}
Thickness fluxes exceed relative vorticity fluxes by an order of magnitude.  Uncertainty metrics represent 90\% confidence intervals.  These estimates are based upon available data from 6 of the 9 LDE moorings with record lengths of approximately 225 days, for a total of 1303 data days.  From a subset of 4 moorings with 434 day records representing 1736 total data days, 
\begin{eqnarray}
\langle {\bf U}^{\prime}\eta^{\prime} \rangle & = & (-0.09 \pm 0.56 ,\;  -0.82 \pm 0.57 ) \times10^{-7} ~{\rm m~s}^{-2}\nonumber\\
\langle {\bf U}^{\prime}\zeta^{\prime} \rangle  & = & (0.86 ,\;  2.69 ) \times10^{-8}  ~{\rm m~s}^{-2} . \nonumber
\end{eqnarray}
We will use both estimates of thickness flux as metrics of the potential vorticity flux below.  

\begin{figure}
\includegraphics[width=0.75\textwidth]{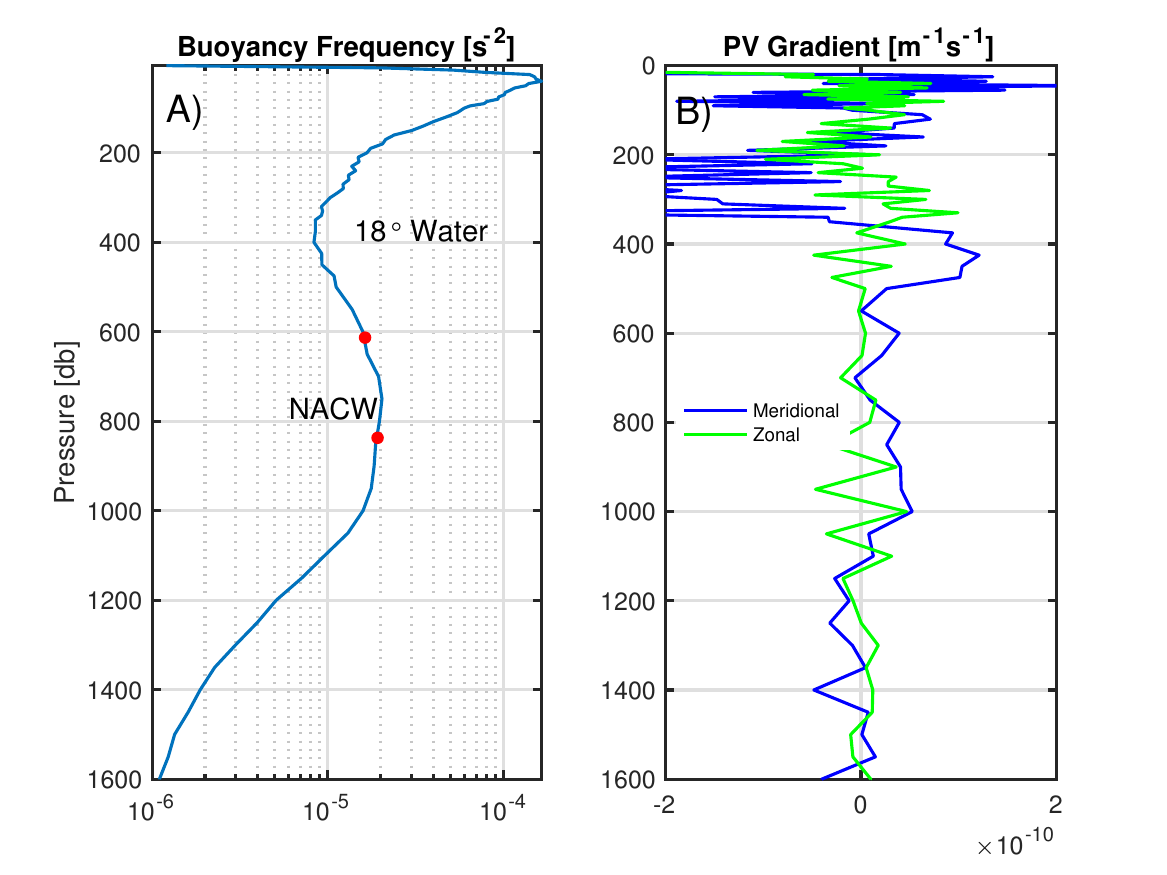}
\caption[]{  Estimates of (A) buoyancy gradient $N^2$ and (B) meridional (blue) / zonal (green) quasigeostrophic potential vorticity gradients (\ref{BackgroundGradients}).  The internal wave - mesoscale eddy energy exchanges estimates come from current meters nominally located at nominal 600 and 825 m depths (dots in panel A), the potential vorticity flux estimates are obtained from the shallower current meters and target 637 m depth.  Note that the zonal gradients in background potential vorticity are smaller than meridional gradients.  The reversal in sign of the meridional gradient at 400 m depth is associated with Eighteen Degree Water and the current meters are located at a depth containing higher stratification associated with North Atlantic Central Water.   }\label{fig:PVgradients}
\end{figure}

Potential enstrophy production is the dot product of the eddy flux and the background potential vorticity gradient.      
A map of planetary vorticity on the potential density $\sigma_{\theta}=27.0$ surface in \cite{Robbins} implies $\nabla_h \langle Q \rangle \cong (0, \beta) = (0, 2\times 10^{-11}$ m$^{-1}$ s$^{-1}$) at the level of the current meter data (the density surface is at approximately 700 m and the current meters are located at approximately 637 m for these potential vorticity flux estimates). We obtain similar estimates of the background potential vorticity gradient from the ARGO climatology of \cite{wijffels2024mesoscale} using geostrophy and the hydrostatic relation, Figure \ref{fig:PVgradients}:
\begin{eqnarray} \label{BackgroundGradients}
\nabla_h \langle Q \rangle & = & \big[ f_0^2\frac{\partial}{\partial_z}\frac{\langle V_z \rangle}{N^2} \; , \; \beta-f_0^2\frac{\partial}{\partial_z}\frac{\langle U_z \rangle}{\langle N^2 \rangle} \big] \nonumber \\
& \cong & \big[\;\;\;\;\;\;\;\;\;\; 0 \;\;\;\;\;\; , \; 2\times10^{-11} {\rm m}^{-1}{\rm s}^{-1}\big]
\end{eqnarray}

Thus, 
\begin{equation}
\langle {\bf U}^{\prime} Q^{\prime} \rangle \cdot \nabla_h \langle Q \rangle \cong -2.9\times10^{-18} ~{\rm s^{-3}}~(1303 \; {\rm days}) \;\;\;{\rm or} \;\;\; -1.6\times10^{-18} ~{\rm s^{-3}}~(1736 \; {\rm days}) . \nonumber 
\end{equation}
\subsubsection{The Enstrophy Cascade}
Enstrophy production occurs at the horizontal scales characterizing potential vorticity fluxes, which turn out to be significantly larger than those characterizing potential enstrophy dissipation.    

The LDE array was specifically designed to estimate terms in the quasigeostrophic potential vorticity equation with mooring spacing to optimally sample the deformation scale horizontal velocity gradients.  The right-hand-side of the potential enstrophy balance, though, concerns second derivatives of horizontal velocities that are dominated by significantly smaller horizontal and vertical length scales than give rise to the surprisingly robust estimate of the wave stress - deformation rate of strain covariances in figure \ref{diff_h}, i.e. $L=100$ km, which in turn is greater than the deformation radius associated with the first baroclinic mode, $L_d = \frac{N}{f}H = 40$ km, using the height scale $H$ identified earlier.  A direct  estimate of potential enstrophy at the dissipation scale is not observationally available and this gradient variance requires some sleuthing.  

To estimate the horizontal eddy spectrum at dissipation scales, we assume that potential enstrophy is cascaded to smaller horizontal scales through nonlinear eddy-eddy interactions.  This cascade across spatial scales is effected by triple correlations in the nonlinear transport term $\mathcal NL$.  Note that the spatial transport, but not the downscale tranfer of potential enstrophy, vanishes in a spatially homogeneous system.  We base this estimate in surface geostrophic velocity estimates from AVISO.  We take one snap shot of that field over a 10 degree by 10 degree area centered on the LDE array to create a horizontal wavenumber spectrum of horizontal velocity.  This is taken as a spectral shape whose amplitude is normalized to the sub-inertial velocity variance of the LDE currents at 600 m, 0.017 m$^2$ s$^{-2}$.  

\begin{figure}
\includegraphics[width=0.75\textwidth]{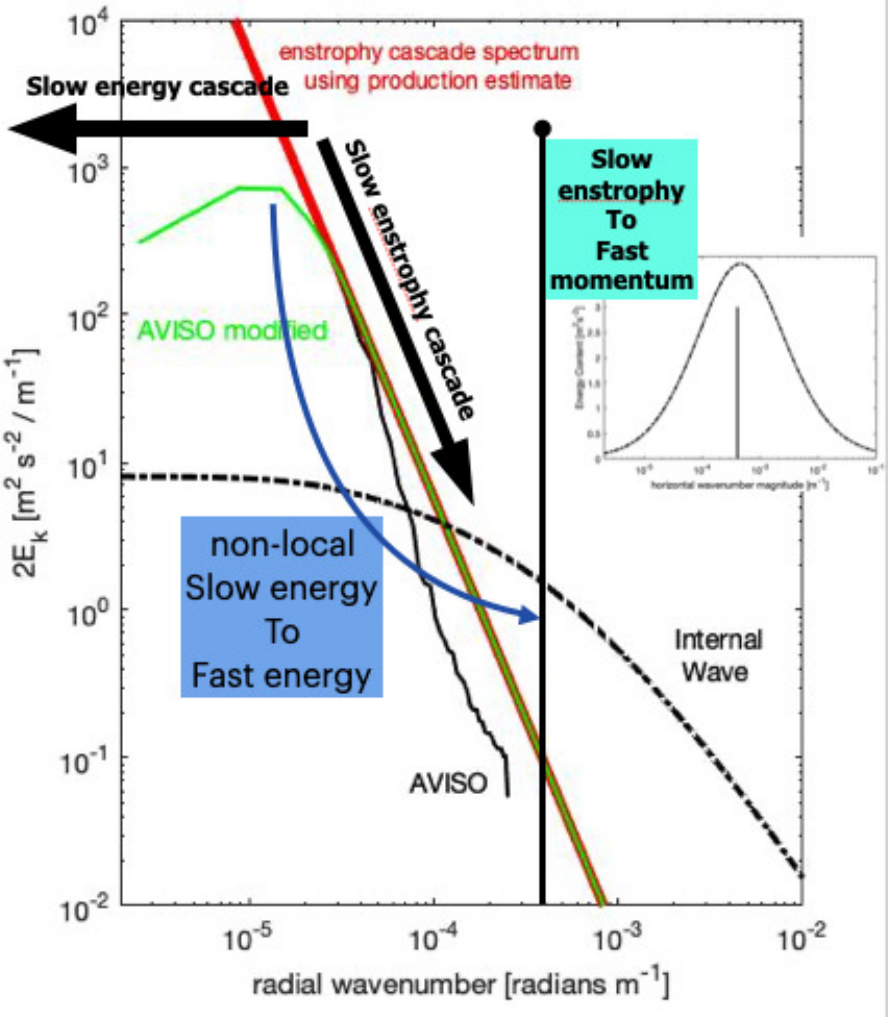}
\caption[]{  (black solid trace) Horizontal wavenumber spectrum of geostrophic velocity from AVISO.  (red trace) Horizontal wavenumber spectrum for the enstrophy cascade regime using the production estimate to set the downscale transfer rate $\gamma = 2.9\times10^{-18}$ s$^{-3}$ s.  (green trace) Horizontal wavenumber spectrum of geostrophic velocity from AVISO, modified within the enstrophy cascade regime.  These three traces are overlie each other.  (black dashed trace) The regional internal wave spectrum. Slow manifold energy and enstrophy cascade paradigms are depicted with bold arrows.  (blue arrow) Transfer of energy from the mesoscale eddy field to the internal wavefield through extreme scale separated interactions is depicted with a smaller weighting.  We argue that the potential enstrophy cascade, which nominally occurs without energy transfers, ends at the energy containing scales of the internal wavefield as submesoscale eddy potential vorticity is traded for internal wave pseudomomentum.  This internal wave energy containing scale is located by the sign post.  See the inset for the variance preserving spectrum. }\label{AVISOCorrelation}
\end{figure}

The AVISO product is understood to have similar constraints on horizontal resolution as the LDE moored array, e.g. \cite{arbic2013eddy}, with the consequence that the AVISO estimate of enstrophy dissipation is biased low.  To address this limitation we consider that dimensional analysis of the enstrophy cascade regime \citep{charney1971geostrophic} leads to 
\begin{equation}
E_1(K_h) = A \; \gamma^{2/3} \; K_h^{-3}
\end{equation}
in which $\gamma$ is the rate at which potential enstrophy is transferred to smaller scales.  We take $A=2.9$ from a numerical study of stratified quasi-geostrophic turbulence \citep{vallgren2010charney} and find, after setting the downscale transfer rate $\gamma$ to be equal to the estimated potential enstrophy production rate, $\gamma = 2.9-1.6\times10^{-18}$ s$^{-3}$, that the enstrophy cascade regime intersects the AVISO spectrum at the deformation scale, figure \ref{AVISOCorrelation}.  These modifications to the AVISO spectrum are pleasing to the eye.  However, the low wavenumber portion of the spectrum is quantitatively irrelevant and thus the fact the low wavenumber portion is arrived by an single realization of the AVISO product from January 1, 2011 is immaterial.

\subsubsection{Locating Enstrophy Dissipation in $K_h$ }\label{ScaleDependent}
In Section \ref{Coupling} our efforts were focused on the transfer of energy.  The use of a single length scale $L$ is tenable within the context of an extreme scale separated limit for energetics:  internal wave - mesoscale eddy energy transfer is proportional to the gradient spectra which in turn are proportional to $K_h^{-1}$ in the potential enstrophy cascade regime.  Alternately, wave refraction is controlled by the background velocity gradients,  gradient content spectra are independent of scale and thus the outer scale is taken.  This is conventional wisdom from time immemorial.   The issue of potential enstrophy dissipation is far more nuanced as the nominal potential enstrophy gradient spectra increase in proportion to $K_h^{+1}$ in the potential enstrophy cascade regime and thus the enstrophy dissipation spectrum needs to be weighted by an appropriate scale dependent viscosity $\nu_h(K_h)$.  

\begin{figure}
\includegraphics[width=35pc]{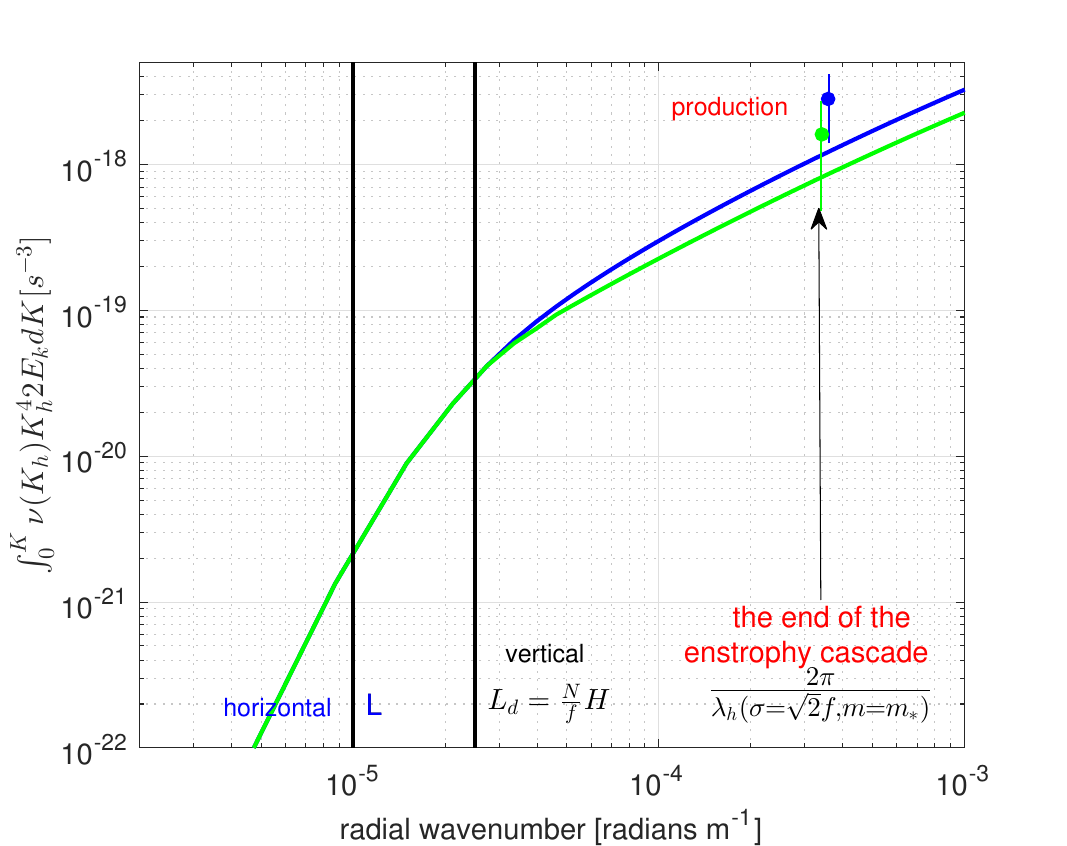}
\caption[]{ Estimates of enstrophy dissipation.  The green and blue lines attempt scale dependent estimates of the potential enstrophy dissipation using 'modified' versions of the AVISO spectrum using observed metrics of the potential enstrophy production rate.  These observed metrics are represented as solid dots with 90\% uncertainty estimates.  These scales are differentiated from the spatial scales characterizing energetics, $L = 100$ km in the horizontal and $L_d= \frac{N}{f} H \cong 40$ km in the vertical.  The `modified' scale dependent dissipation estimate is approximately equal to half the production rate at the horizontal wavelength of the energy containing scale of the internal wave horizontal spectrum, coinciding with $k_h(\sigma =\sqrt{2}f, m=m_{\ast}) = j_{\ast}/L_d$.  This approximate equality is interpreted as, `The End of the Enstrophy Cascade'. }\label{EnstrophyProductionDissipation}
\end{figure}

The issue is that the math envisioned around (\ref{BadBoy}) does not allow this to be explicit.  Thus, an {\it ad hoc} approach of allowing for a scale dependence through $\tau_p(L = K_h^{-1})$ is taken.  Representing enstrophy dissipation in the spectral domain, we argue that 
\begin{eqnarray}\label{EnstrophyDissipation}
\frac{1}{2} \nu_h[\langle \zeta_x^{\prime 2}  + \zeta_y^{\prime 2} \rangle + \frac{f_o^2}{ \langle B_z \rangle } \langle \zeta_z^{\prime 2} \rangle ] 
& \rightarrow & \int_0^{K_h} 2 \nu_h(K_h^{\prime}) \; K_h^{\prime 4} \; E_1(K_h^{\prime}) \; dK_h^{\prime} 
\end{eqnarray}
with $\nu_h(K_h)$ given by (\ref{visc}) and $\tau_p^{-1}=C_g^h K_h$.  The factor $E_1(K_h)$ represents the 1-d horizontal spectrum of eddy kinetic energy.  Empirical scalings of $\zeta^2 \cong 2K_h^2E_1(K_h)$ and $\frac{f_o^2}{\langle B_z \rangle }\partial_z^2 \cong \nabla_h^2 $ have been invoked to arrive at the prefactor in (\ref{EnstrophyDissipation}).  
Evaluation of the potential enstrophy dissipation function (\ref{EnstrophyDissipation}), Figure \ref{EnstrophyProductionDissipation}, returns a potential enstrophy dissipation rate estimate of half the potential enstrophy production rate at the horizontal length scales characterizing the energy containing near-inertial internal waves, $K_h = 2\pi / \lambda_h(\sigma=\sqrt{2}f, m=m_{\ast})$!

\subsubsection{Physical Interpretation of the Enstrophy Dissipation Scale }
Our intuition is that the physics of trading eddy potential enstrophy for internal wave momentum is arrived at the envelope scale.  Thus our expectation is that the spectral dissipation (\ref{EnstrophyDissipation}) should be similar to the downscale enstrophy transfer rate $\gamma$ at the envelope scale $\mathcal L$ of the energy containing near-inertial waves.  Observational metrics of such packet structure are minimal.  Physical intuition suggests a relatively large bandwith $\Delta \p$ associated with the envelope, e.g. $\Delta \p \cong \p$, so that a Fourier uncertainty principle 
\begin{eqnarray}
    \Delta \r \cdot \Delta \p & \ge & 1/2 \nonumber \\
    & {\rm suggests}& \\
    \mathcal{L} \; k_h & \overset{>}{\sim} & 1/2 \nonumber
\end{eqnarray}
Thus our expectation is that the eddy enstrophy dissipation scale should be similar to the energy containing scale of the near-inertial field, i.e. similar to $2 \pi / \lambda_h(\omega=\sqrt{2}f,m=m_{\ast})$. 

\subsubsection{Double Checking}
\begin{figure}
\includegraphics[width=35pc]{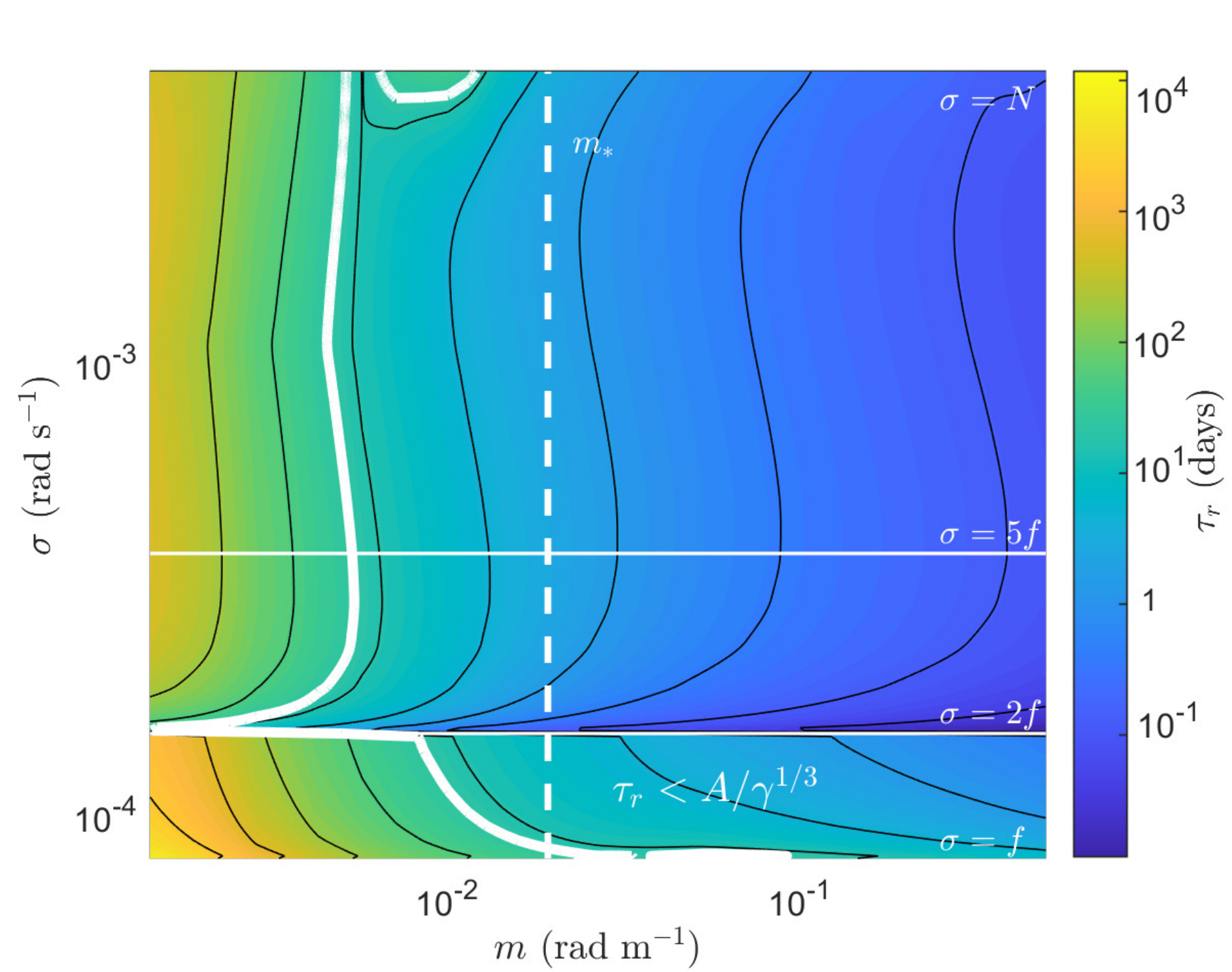}
\caption[]{  A contour plot of horizontal nonlinear relaxation time scales (figure \ref{GiovanniTimeScale}) with the enstrophy cascade time scale $A/\gamma^{1/3}$ represented by the thick white contour and the energy containing scale of the vertical wavenumber spectrum represented as the dashed white vertical line.  The nonlinear relaxation time scale is shorter than the enstrophy cascade time scale to the right of the white contour.  An approximate equivalence of these time scales within the energy containing part of the near-inertial field ($m_{\ast}$) is interpreted as indicating a dynamical linkage of the fast (gravity dominated) and slow (rotationally dominated) manifolds.  }\label{EnstrophyRelaxationTimescale}
\end{figure}

In order to demonstrate that this quite remarkable result is not simply the result of one {\em ad hoc.} assumption piled upon another, we perform a consistency check.  We directly compare the nonlinear relaxation timescale with that characterizing the potential enstrophy cascade, $A/\gamma^{1/3}$, in figure \ref{EnstrophyRelaxationTimescale}.  We find that the time scale for nonlinear interactions to relax perturbations at the energy containing scales of the near-inertial field is shorter than the downscale enstrophy transfer rate.  

\subsubsection{The Vertical Coordinate}
Formulating a corresponding scale dependent estimate of the vertical viscosity is more complex.  Halving the height scale $H$ decreases the propagation time scale $\tau_p$, which in turn significantly reduces the positive contributions to the effective vertical viscosity coming from relatively large vertical wavelengths, figure \ref{integrand_v}.  This tendency could be offset by minor frequency dependent changes to the vertical bandwith $m_{\ast}$ and use of a non-hydrostatic dispersion relation.  Observational constraints here are minimal.  


\label{Enstrophy_Budget}

\section{Summary and Discussion}\label{Summary} 
\subsection{Summary}\label{Results}

A modified version of the mesoscale eddy--internal wave coupling mechanism described by \cite{M76} is used to predict an effective horizontal viscosity of 50 m$^2$ s$^{-1}$ and an effective vertical viscosity of $2.5\times10^{-3}$ m$^2$ s$^{-1}$ .  The model requires specifying a stationary isotropic homogeneous background spectrum, which is then perturbed through mesoscale eddy interactions.  Subtle regional variability in background spectra are documented in \cite{polzin2011toward}.  This includes high wavenumber - high frequency power laws, vertical wavenumber bandwidth and the inertial peak.  The model - data comparison presented here requires engaging with this variability:  a specific chracterization of background internal wavefield was created for the Sargasso Sea.  Having done so, the modified theory is in remarkable agreement with energy exchanges between waves and eddies at the energy containing scales of the eddy field in the Local Dynamics Experiment. 

The LDE was a large multi-investigator program resulting in a broad swath of work on eddy energetics and eddy dynamics \citep{kamenkovich1986polymode}.  This body of work allows us to enquire rather than speculate about the consequences.  The most significant result is a statistically significant estimate of potential vorticity fluxes directed across mean potential vorticity gradients.  This implies potential enstrophy production with an attendant cascade of potential enstrophy to smaller scales.  We have estimated a length scale dependent wave-eddy coupling, and our work implies the transfer of eddy potential vorticity to internal wave pseudomomentum at the energy containing scale of the internal wavefield.  This remarkable result provides a dynamical rationale for locating the end of the enstrophy cascade at the energy containing scale of the near-inertial field.  These dynamics likely dictate the regional $m_{\ast}$, whereas energetics determine the spectral density $B m_{\ast}$ and high wavenumber cutoff $m_c$ through the connection of the internal wave vertical wavenumber gradient spectra to turbulent dissipation $\epsilon$.

\subsection{The Pathway to Theoretical Upgrades}\label{Optics}
\subsubsection{A First Principles Derivation of the Boltzmann Analogue}\label{FirstPrinciples}
A point of tension in our assessment of potential enstrophy dissipation is the implementation of a scale dependent viscosity operator $\nu_h(\tau_p^{-1}= C_g^h K_h)$ to effectively filter the potential enstrophy dissipation spectrum.  A first principles derivation would require the following steps.  

At (\ref{SourceTerm}) we note that 
\begin{equation}
(\dot{k}, \dot{l} ) \cdot \nabla_{(k,l)} \rightarrow (k \dot{k} , l\dot{l} ) k_h^{-1} \nabla_{k_h} \; .\
\nonumber
\end{equation}
Since the evolution of wavenumber is being evaluated along ray paths, there is a direct analogy with a Lagrangian analysis presented in \cite{taylor1922diffusion}, \citep{polzin2017oceanic}.  For expository purposes we replace $k$ with $\psi$:
\begin{eqnarray}
    \dot{k}(t) \; k(t) \rightarrow \dot{\psi}(t)[\psi-\psi(t=0)] & \equiv & \int_{0}^t \dot{\psi}(t) \; \dot{\psi}(t^{\prime}) \; dt^{\prime} \nonumber \\
    \frac{1}{2} \frac{d}{dt} [\psi-\psi(t=0)]^2 & \equiv & \int_0^{t} \dot{\psi}(t) \; \dot{\psi}(t-\tau) d\tau \end{eqnarray}
with $t$ representing time along a ray.  The phrasing from the Lagrangian analysis is that, {\em IF} the integral converges, the integral defines a diffusivity $\mathcal K$
\begin{equation}
{\mathcal K} = \langle \dot{\psi}^2 \rangle \tau_c
\nonumber
\end{equation}
where $\langle \dot{\psi}^2 \rangle$ is the analog of a velocity variance and $\tau_c$ is the lagged autocorrelation time scale.  The issue here is that while one can envision wave refraction as a spatially homogeneous problem, wave refraction is distinctly inhomogeneous in wavenumber space and the integral will not converge \citep{polzin2017oceanic,lvov2024generalized,polzin2025one} in association with a mean drift $\langle {\dot p} \rangle \ne 0$ in wavenumber.  

The convergence issue is somewhat mollified by correcting for the mean drift, i.e. 
\begin{equation}
\int_{0}^{t} \langle \dot{\psi}(t)\dot{\psi}(t-\tau) \rangle d\tau \rightarrow \int_{0}^{t}\langle \big[\dot{\psi}(t)-\langle \dot{\psi}(t)\rangle\big]\big[\dot{\psi}(t-\tau))-\langle \dot{\psi}(t-\tau)\rangle \big] \rangle d\tau + {\rm mean \; drift}
    \label{Taylor21}
\end{equation}
This will produce an advective action transport represented above as a 'mean drift' in addition to the diffusive transport term.  The time scale associated with the mean drift, $p/\langle {\dot p} \rangle$, has further utility.  In comparison to the correlation time scale $\tau_c$, the limit $\tau_c < p/\langle {\dot p} \rangle$ defines approximate convergence of the integral in (\ref{Taylor21}).  This limit can be identified as a Markov process.  

Information about the background velocity gradient spectrum is available through the Wiener-Khinchin theorem:
\begin{equation}\label{eq:handwave}
    \int_{0}^{t} \langle \dot{\psi}(t)\dot{\psi}(t-\tau) \rangle d\tau  = \int_0^{t} \int_{-\infty}^{\infty} [P_{{\dot \psi}{\dot \psi}}(\omega) e^{i \omega \tau} d\omega \big] d\tau
\end{equation}
where $P_{{\dot \psi}{\dot \psi}}$ is the power spectral density of $\dot \psi$, i.e. (\ref{eq:Eikonal}), in time $\tau$ along the ray path.  This ultimately relates to the background spatial structure through the background phase $Kx+Ly+Mz-\Omega t = \displaystyle \int ({\bf C}_g + {\bf U} )d\tau -\Omega t$.  Fourier transforming (\ref{eq:handwave}), as indicated below (\ref{eq:NearlyThere}), returns a filter through the $\displaystyle \int_0^{t} d\tau$ operation rather than the {\em ad hoc} invocation in Section \ref{budgets}.\ref{Potential_Enstrophy}.\ref{ScaleDependent}.  Development of this first principles approach is beyond the scope of this manuscript.  

The invocation of a Fourier transform in (\ref{PropagationTimeScale}) brought forward the issue of a phase velocity - group velocity resonance, but that resonance is avoided in order to develop the issue of damping.  The resonance will map onto a mean drift $\langle \dot{p} \rangle$ in wavenumber space.  This mean drift implies potential enstrophy dissipation that complements the diffusive (viscous) contribution identified in the prior section.

\subsubsection{Including $n_3^{(2)}$}\label{WKB}
The proposed enstrophy dissipation scenario is one in which potential vorticity on the packet scale is traded for momentum flux gradients captured by $n^{(2)}$.  We do not address $n^{(2)}$ directly.  Rather, we judge the efficacy of enstrophy transfers using a flux-gradient closure for the perturbation action spectrum $n^{(1)}$.  We believe this to be an appropriate order of magnitude characterization for the following reason.

This closure is grounded in the lowest order of a scale separation between the waves and the background flow, with ray trajectories estimated from variations in the Doppler shift.  In the context of a formal WKB asymptotic expansion of a wave equation, the first term represents the amplitude, the second phase, and succeeding contributions come in ordered amplitude-phase contributions \citep{bender2013advanced}.  Additionally, within each pair, the phase contribution is not asymptotically smaller than the amplitude.  The first term in this expansion is referred to as the Geometric Optics approximation, the second term as Physical Optics.  We understand ray tracing with action spectral density being conserved along ray trajectories in  phase space using only the Doppler shift as representing the Geometric Optics approximation.  

The formal WKB analysis of a 3-D wave equation has not been done, due to the difficulty in deriving the 3-D equation \citep{Polzin96a}.  Rather, what we have \citep{young1997propagation} are expressions for a dispersion relation invoking a near-inertial limit in which Doppler shifting is assumed to be small and one relaxes the extreme scale separation to incorporate background gradients of $O({\ell}/{\mathcal L})$.  The deformation rate of strain does not enter these dispersion relations.   Rather, the lower bound of the internal waveguide is modified from $\omega = f$ to be proportional to the potential vorticity \citep[]{hoskins1974role}.  We identify this as being consistent with the Physical Optics approximation of a formal WKB expansion, in which Geometric Optics typically represents the controlling factor and Physical Optics represents a correction, \cite{bender2013advanced}.  

We take this characterization as support for our identification of the end of the enstrophy cascade.  Substantial algebra will be required to update the polarization relations (\ref{eq:PolarizationRelations}) and expressions to the rate of refraction (\ref{Characteristics.org}) $\dot{\r}$ and $\dot{\p}$, to systematically include the $O({\ell}/{\mathcal L})$ terms.  

\subsection{Implications}\label{Interpretation}
\subsubsection{Mesoscale Dynamics}
Potential vorticity is conserved in the absence of diabatic processes and friction - \citep{Ertel, HMcI}.  The dominant intellectual prejudice in Physical Oceanography is to regard frictional processes as being essentially diabatic in nature and sufficiently weak within the oceanic interior that potential vorticity modification following a parcel occurs only at the boundaries.  Within an eddy only paradigm, we have a robust understanding of nonlinear transfers from baroclinic to barotropic eddies and an upscale (inverse) energy cascade coinciding with a downscale (forward) potential enstrophy cascade. \citep[e.g][]{charney1971geostrophic,Rhines79,Salmon}. Our prognostic model indicates that internal waves extract energy from the mesoscale eddies at the horizontal and vertical scales that characterize baroclinic instability and potential vorticity fluxes, complicating this eddy only paradigm.  

Our findings imply that enstrophy dissipation is of similar order of magnitude to enstrophy production in the one observation based data set for which such an estimate is possible.  We suspect that this is not coincidental as the potential vorticity - wave momentum exchanges provide a rational for determining the vertical wavenumber bandwidth $m_{\ast}$.  However, this similitude {\em does not} preclude the appearance of spatially localized regions of upgradient PV fluxes and / or spatially localized regions of significant potential enstrophy transport divergence \citep{marshall1981note} that are a part of recirculating jet problems \citep{waterman2011eddy}.  While the setting of the LDE data is the Southern Recirculation Gyre of the Gulf Stream, the background PV gradient, figure \ref{fig:PVgradients}, is that of zonally oriented watermass structures rather than a jet entry - jet exit paradigm and we are happy to take the evidence \citep{BOB86} of down gradient potential vorticity fluxes at face value.  On the other hand, we note that idealized models in an inviscid limit exhibit a PV homogenization scenario \citep{RY82,WW04} in which mesoscale eddies transport potential enstrophy to the boundaries and destroy the waveguide that planetary waves are riding on.  Our considered opinion is that the ocean likely does not lie in that part of this PV homogenization parameter space.  Insight into the recirculating jet problem could be gained by constructing the eddy-mean analogue of this wave-eddy problem, in which baroclinic instability \citep{eady1949long} represents a source, the physics of an inverse energy cascade and upgradient potential vorticity fluxes \citep[e.g.][]{Starr} are obtained by replacing the internal wave polarization relations (\ref{eq:PolarizationRelations}) with the thermal wind relation and the relaxation timescale is represented as a combination of internal wave coupling and eddy-eddy nonlinearity.  The \cite{marshall2012framework} framework, in which internal wave coupling would appear as 'PV mixing' is highly instructive in this context.  

These ideas and concepts are ripe for exploration with the availability of expanding climatologies from Argo \citep{whalen2018large, pollmann2020global, wijffels2024resolving, hersh2025long}, moored current meter \citep{le2021variability, dematteis2024interacting} and EM-profiling \citep{girtonSquid} data bases.  These ideas and concepts are ripe for exploration with comparisons of regional characterizations of the internal wavefield with those from high resolution regional models \citep{pan2020numerical, nelson2020improved, barkan2021oceanic, yang2023oceanic, delpech2024eddy, barkan2024eddy} and the internal workings of those models \citep{skitka2024internal}.  
 
\subsubsection{Climate}
One perspective of climate change is that temporal trends at increasingly longer time scales represent increasingly smaller imbalances in the equations of motion.  If true, can we really claim to understand how the climate system works if GCMs are simply tuned to today's conditions and do not address the proper sub-grid scale physics?  What happens to the behavior of the general circulation as the thermocline tightens, the eddy scale decreases, or the rms eddy  velocity increases?   What are the phenomenological linkages between internal waves and eddies; and the consequences for eddy dynamics?

\clearpage
\newpage

\clearpage
\acknowledgments
Much of the intellectual content of this paper evolved out of discussions with R. Ferrari several decades ago.  The manuscript benefited from similarly dated discussions with Brian Arbic, Rob Scott and a review provided by Peter Rhines.  Salary support for this analysis was provided by NSF through OCE-2319144 and OCE-2232439 (NOPP), and the Henry M. Stommel Chair in Oceanography. GD acknowledges funding by the Simons Collaboration on Wave Turbulence (Award No. 652354) and by the European Union (ERC, Project Acronym: OPPIWaM, Project No.: 101222068). An early version of this work was submitted to JPO in 2009 and returned without review by the Chief Editor, with the opinion that the substantive dynamical issue was in the degradation of the near-inertial waveguide by reversals in the sign of potential vorticity.  This opinion is  inconsistent with the quasi-geostrophic setting of the LDE.  


%
%
\datastatement
No data were created for this paper.  Just reruns of the classics.  

%

\appendix[A]\appendixtitle{Other Assessments}\label{Others}
\subsection{Observations}

Observational assessments of internal wave - mesoscale eddy energy exchanges from moored arrays are provided in \cite{cusack2020observed} (the Drake Passage as part of DIMES), \cite{savage2025observations} (the NE Atlantic as part of OSMOSIS) and \cite{jing2018observed} (the Gulf of Mexico).  These lack quality vertical wavenumber domain information to ground the analysis pursued here.  

\subsection{Theory}
\subsubsection{\cite{M76}}
The Sargasso Sea observations and our theoretical estimates of the vertical exchange coefficient are both much smaller than M\"uller's prediction of $\nu_v + \frac{f^2}{N^2} K_h \cong 0.45$ m$^2$ s$^{-1}$, which was originally repudiated by observations presented in \cite{Frank76} and \cite{RJ79}, and is more than two oders of magnitude larger than that provided in \cite{polzin2010mesoscale}.  

M\"uller's large estimate is a product of (at least) three compounding issues:
\begin{enumerate}
\item Use of a non-hydrostatic dispersion relation, which appears to be inconsistent with a vertical modal structure \citep{pinkel1975upper} at large vertical scales and potential linkage to super-buoyancy, transient nonlinear interactions at smaller vertical scales.  
\item Assuming the eddy interactions were local in space/time, i.e. $\tau_r \ll \tau_p$
\item Characterizing the relaxation time as being independent of vertical wavenumber and frequency, even though the Bragg scattering process was identified as the crucial relaxation physics.  
\end{enumerate}

\subsubsection{\cite{RJ79} }
\cite{RJ79} note that \cite{M76}'s zeroth order wavefield is specified as the isotropic universal (GM) model in an Eulerian frequency coordinate.  They argue that the relaxation process could very well be toward an equilibrium spectrum with an intrinsic frequency coordinate.  This appears to have some parallels with our Introduction.  Using the 'Induced Diffusion' piece of the resonant manifold, they obtain a solution to the radiation balance equation that is a combination of thermodynamic equilibrium $n_3^{(0)}(\p) \propto k_h/ \mid m \mid$ and non-equilibrium no-flux $n_3^{(0)}(\p) \propto k_h^{-x}m^{0}$ solutions.  There are decided issues with the 'Induced Diffusion' mechanism that are discussed at length in \cite{lvov2024generalized} and \cite{polzin2025one}.  Note that the Sargasso Sea spectrum, which we promote as representing an wave-eddy end-member of forcing space, is inconsistent with $n_3^{(0)}(\p) \propto k_h^{-x}m^{0}$.  

\subsubsection{\cite{W85} }
The genesis of \cite{W85} is a recognition that there is no mechanism in the resonant interaction scheme of \cite{MMb} for transporting energy within the high vertical wavenumber ($100 \geq \lambda_v \geq 10$ m) near-inertial ($f < \omega \leq 2f$) frequency band toward even higher vertical wavenumber.  The intent of \cite{W85} was to formulate a radiation balance representation of wave--mean interactions in this band to transport action to a sink at $\lambda_v < 10$ m through the cumulative effects of near-inertial wave refraction in the mesoscale fields.  Our understanding is that near-inertial energy at these vertical wavelengths is transported to higher frequency in association with a Bragg scattering branch of the resonant manifold, figures \ref{GiovanniTimeScale} and \ref{TimeScalesTFN}.  To reiterate:  our focus here is analogous to the development of a surface wind wave field rather than the propagation of surface swell.  

\subsubsection{Subsequent undamped models }
We reprise:
\begin{quote}
     Energy exchange can be accomplished if either the forcing is in phase with the damping or if the forcing is in resonance.
\end{quote}
The pioneering work of \cite{W85} is more recently pursued in many GFD oriented publications.  These break down along a short time behavior within a $C_{{\rm g}} \gg U$ limit that exhibits no energy exchange with the mesoscale \citep{savva2018scattering, kafiabad2019diffusion, savva2021inertia, cox2023inertia} and a long time behavior in a $C_{{\rm g}} \ll U$ limit that supports energy exchange \citep{dong2020frequency,  dong2023geostrophic}.  The short time behavior is one in which the mesoscale velocity $U$ is characterized as a frozen field.  The long time behavior that supports energy exchange is also attended by waves with smaller spatial scales and thus a greater impact of time dependence in the mesoscale.  We find that the wholesale neglect of ``damping'' in the Boltzman analog limits the applicability of these ideas to our work, in which the relaxation time scales are typically equal to or shorter than propagation time scales, Figure \ref{TimeScalesTFN}.  



A key objective for such refractive models is the identification of a diffusivity tensor $D_{ij}$ in wavenumber space:
\begin{equation}
D_{ij}(\p) = \int_{t-\tau}^{t} \langle \dot{\p}_i(\r(t))\dot{\p}_j(\r(t^{\prime})) \rangle dt^{\prime} \rightarrow \tau_c \langle \dot{p_i}\dot{p_j} \rangle 
\label{DiffusivityTensor}
\end{equation}
that hinges upon the time integration converging to a correlation time scale $\tau_c$.  This foundational assumption ignores the importance of resonance conditions that require fundamental modification of transport equations \citep[][Section 3.2.4]{lvov2024generalized} and \cite[][Section 3.b]{polzin2025one}.  Such resonance conditions are highly problematic for internal waves described by the parametric representation (\ref{ParametricSpectralRepresentation}).  The relevance of such resonances is underscored by the oceanographic community's very first look at the 10 meter vertical structure of the horizontal velocity field, \citep{LS75}  being an example of a phase velocity - group velocity resonance \citep{polzin2008mesoscale}.


%
\subsubsection{ The Bounce }
As formulated, these models assume a constant buoyancy frequency.  Waves are free to propagate in the vertical and will terminate an interaction event on a time scale $\tau_p = H/C_g^z$.  Waves of sufficiently high frequency, however, will encounter turning points where their intrinsic frequency approaches that of the local stratification rate $N(z)$.  Curiously, the negative vertical stress - vertical shear correlation occurs for waves that potentially encounter a buoyancy frequency turning point, figure \ref{diff_v}.  The presence of turning points and a boundary can give rise to a variant of the wave capture scenario.  

In a deformation strain horizontal wavenumber magnitude asymptotically increases.  Vertical wavenumber undergoes either an increase or decrease, but at both surface reflection or turning point there will be a sign change and opposing time evolution of the vertical wavenumber.  Over many reflections and turning points, horizontal wavenumber magnitude increases, the vertical wavenumber remains nearly constant and consequently the intrinsic frequency increases.  Such a wave will become progressively trapped in regions of higher and higher buoyancy frequency.  A high frequency wave reported in \cite{JS84} may represent such an event.  It is not clear that this phenomenology is appropriately represented in (\ref{visc-vert}).  The relevance of this mechanism as damping for upper ocean frontogenetic conditions \citep{yu2024intensification} is palpable.


\appendix[B]\appendixtitle{Uncertainty and Sensitivity}\label{UncertaintySensitivity}
This Appendix addresses issues of statistical uncertainty and model parameter sensitivity.  Coherence estimates for the momentum flux estimates are presented in AppendixB.\ref{StatisticalMetrics}.  A brief exploration of model parameters is presented in Appendix B.\ref{ParameterSensitivity1} and B.\ref{ParameterSensitivity2}.  The conclusion to draw from these results is that theoretical efforts need to be fully engaged with Regional parametric representations of the internal wave spectrum.  

\subsection{Statistical Uncertainty}\label{StatisticalMetrics}

This subsection presents estimates of stress-strain and stress-shear coherence that appear as figures 4 and 5 of \cite{polzin2010mesoscale}, respectively.  Statistical metrics are developed by averaging data from multiple moorings each having durations in excess of one year.  In particular, note that integration of the vertical cospectra returns $\nu_v  + \frac{f^2}{N^2}K_h = 2.5 \pm 0.3 \times 10^{-3}$ m$^2$ s$^{-1}$, Figure  \ref{diff_v}.  The error estimate provided here represents the compounding of the nominal 750 degrees of freedom per coherence estimate (Fig. \ref{coh_v}) over a bandwidth of 10-40 cpd containing 320 coherence estimates.  

\begin{figure}[htbp]
\includegraphics[width=35pc]{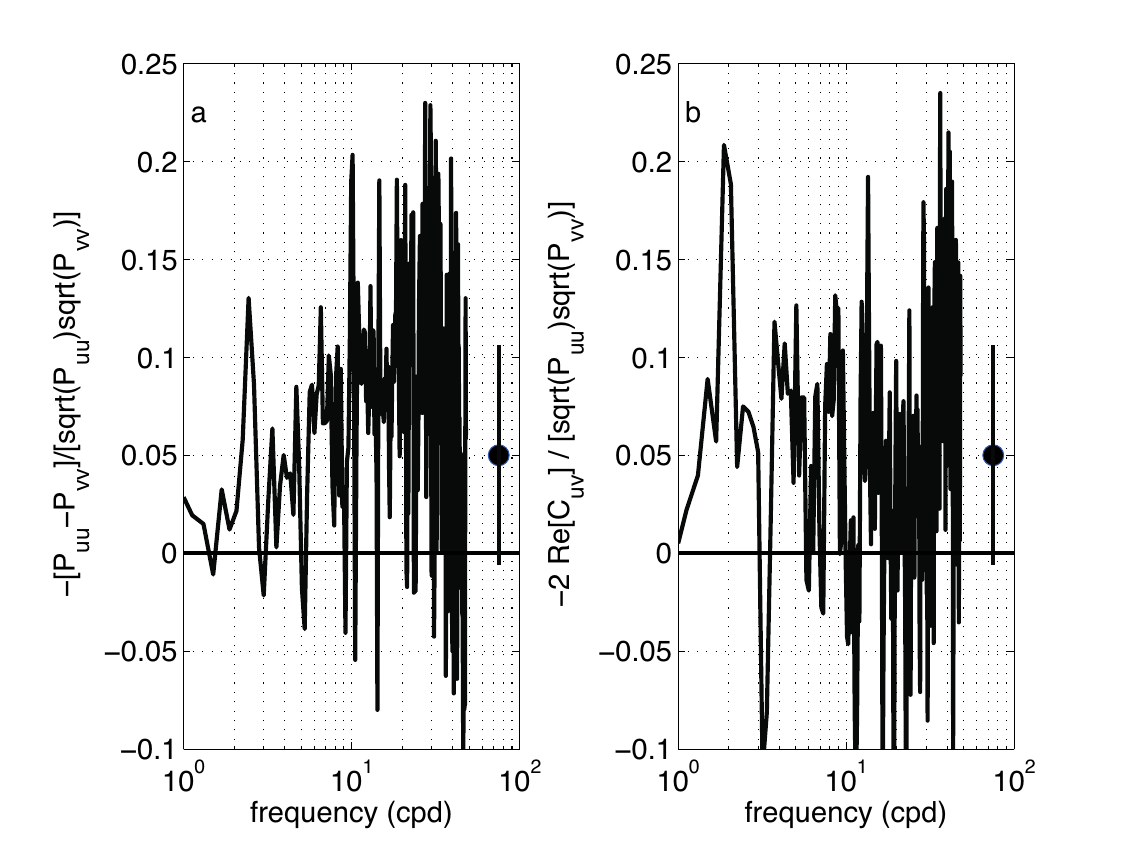}
\caption[]{Coherence functions created by averaging (a) $-sgn(S_n)[P_{uu}-P_{vv}]/P_{uu}^{1/2}P_{vv}^{1/2}$ and (b) $-sgn(S_s) C_{uv} / P_{uu}^{1/2}P_{vv}^{1/2}$ with $C_{uv}$ being the real part of the $uv$ cross-spectrum.  Estimates are based upon 1024 point transform intervals and averaging over two triangles at 825 m water depth.  The symbol on the right hand side of the panels represents the standard deviation at a 0.05 coherence level with the 320 degrees of freedom expected for each cospectral estimate. \label{coh_h}} 
\end{figure}

\begin{figure}[htbp]
\includegraphics[width=35pc]{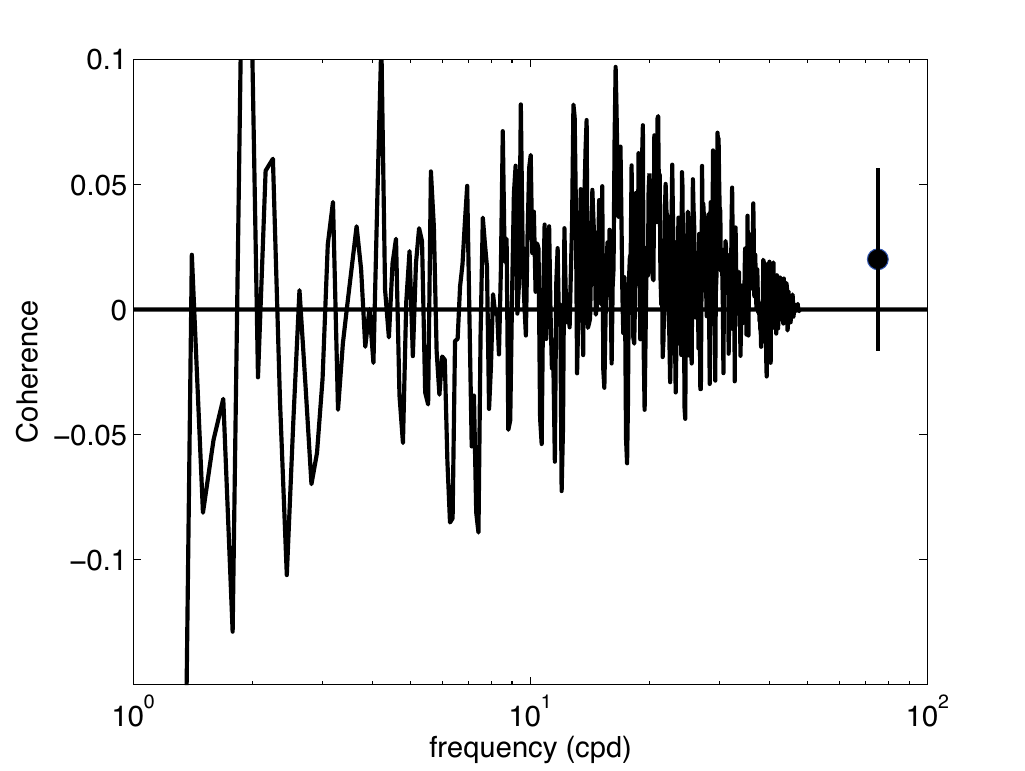}
\caption[]{Coherence function created by averaging $-sgn(U_z)[C_{uw} - fN^{-2}C_{vb}] / T(\omega) P_{uu}^{1/2}P_{ww}^{1/2}$ and $-sgn(V_z)[C_{vw} + fN^{-2}C_{ub}] / T(\omega) P_{vv}^{1/2}P_{ww}^{1/2}$.  The factor  $C_{xy}$ represents the real part of the $xy$ cross-spectrum.  The transfer function $T(\omega) = (\omega^2-f^2)/(\omega^2+f^2)$ accounts for cancelation of the Reynolds stress by the buoyancy flux and renders the denominator consistent with the numerator.  The coherence estimates are based upon 1024 point transform intervals of data at both 600 and 825 m levels.  Data are from the Center, Northeast and Northwest moorings.  The symbol on the right hand side of the figure represents the standard deviation at a 0.02 coherence level with the 750 degrees of freedom expected for each spectral estimate.  \label{coh_v}} 
\end{figure} 

\subsection{Parameter Sensitivity:  Sargasso Sea}\label{ParameterSensitivity1}
\begin{figure}[htbp]
\includegraphics[width=35pc]{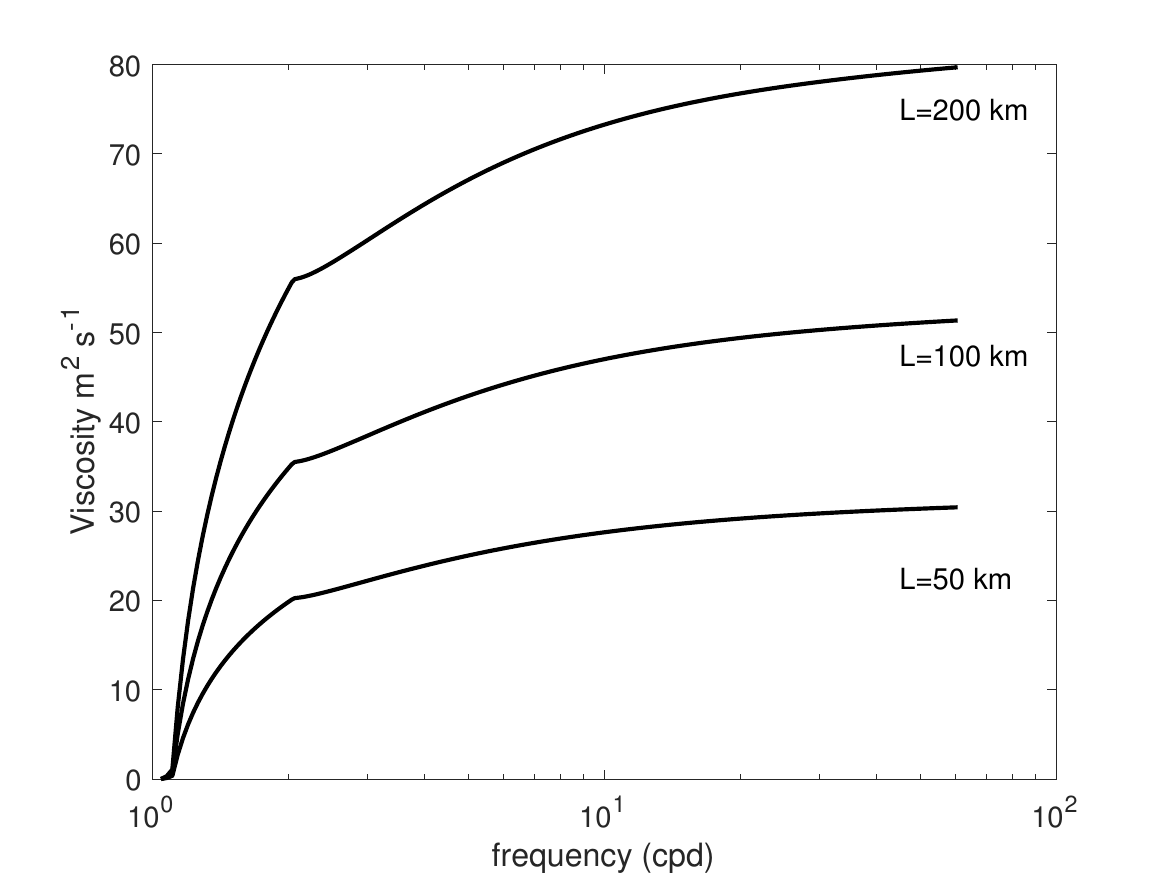}
\caption[]{Sargasso Sea Spectrum derived model estimates of horizontal viscosity.  Black lines represent model estimates of the cumulative frequency integrals for eddy scales $L=[50, 100, 200]$ km using internal wave energy levels (Table \ref{ParametricSpectralParameters}) at 600 m depths.  See figure \ref{diff_h} for the model-data comparison.  \label{model_horz_L} }
\end{figure}

\begin{figure}[htbp]
\includegraphics[width=35pc]{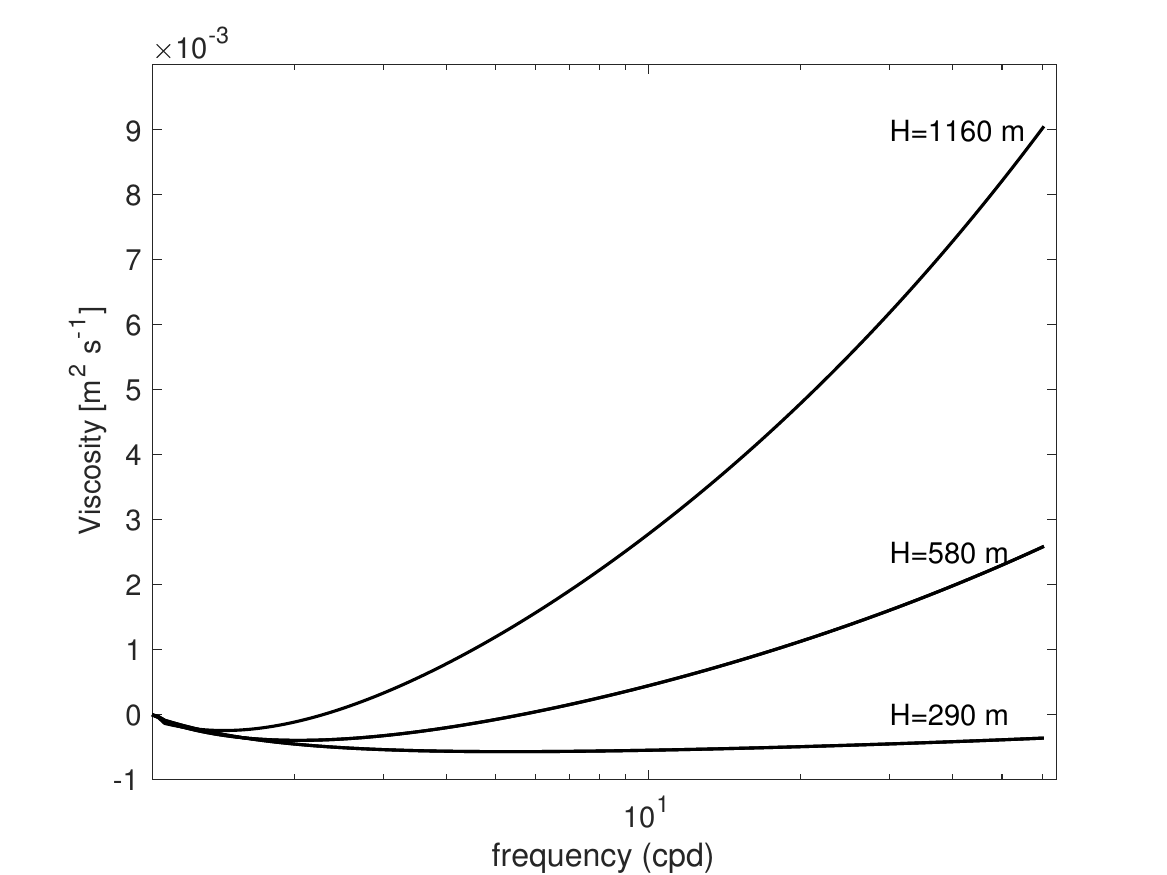}
\caption[]{Sargasso Sea Spectrum derived model estimates of vertical viscosity.  Black lines represent model estimates of the cumulative frequency integrals for eddy height scales $H=[0.5,\; 1.0,\; 2.0]/\pi$ times the buoyancy scale water column height, 1830 m, using internal wave energy levels (Table \ref{ParametricSpectralParameters}) at 600 m depths.  See figure \ref{diff_v} for the model-data comparison. 
}\label{model_vert_H}
\end{figure}

In this subsection we present model estimates of horizontal (figure \ref{model_horz_L}) and vertical (figure \ref{model_vert_H}) viscosity using the Sargasso Sea spectrum utilizing a limited sweep of spatial scales $L$ and $H$ that define the propagation time scale $\tau_p$.  

\subsection{Parameter Sensitivity:  Spectral Parameters}\label{ParameterSensitivity2}

In this section we present horizontal viscosity and effective vertical viscosity estimates for the GM76 model, which is defined by curve fits to wintertime data at Site-D, north of the GulfStream \cite{polzin2011toward}, and a modified version of the Sargasso Sea spectrum that has a vertical bandwidth similar to the GM76 spectrum.  

\begin{figure}[htbp]
\includegraphics[width=35pc]{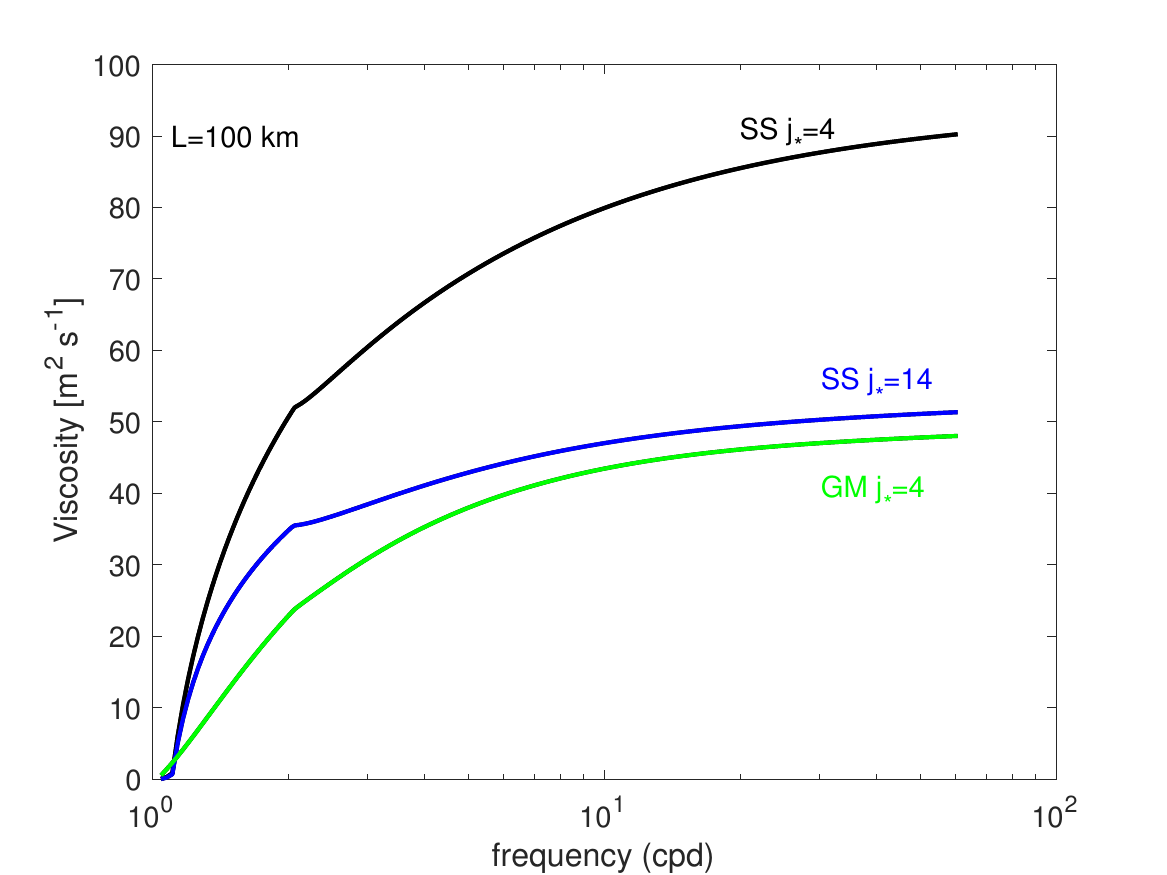}
\caption[]{Model estimates of the horizontal viscosity for an eddy length scale $L=100$ km.  The blue trace represents our reference Sargasso Sea spectrum, the black trace modifies that spectrum for an equivalent mode bandwidth of $j_{\ast}=4$, and the green trace represents the GM76 spectrum, determined by observations at Site-D, north of the GulfStream.  See figure \ref{diff_h} for the model-data comparison.  \label{model_horz_spectrum} }
\end{figure} 

\begin{figure}[htbp]
\includegraphics[width=35pc]{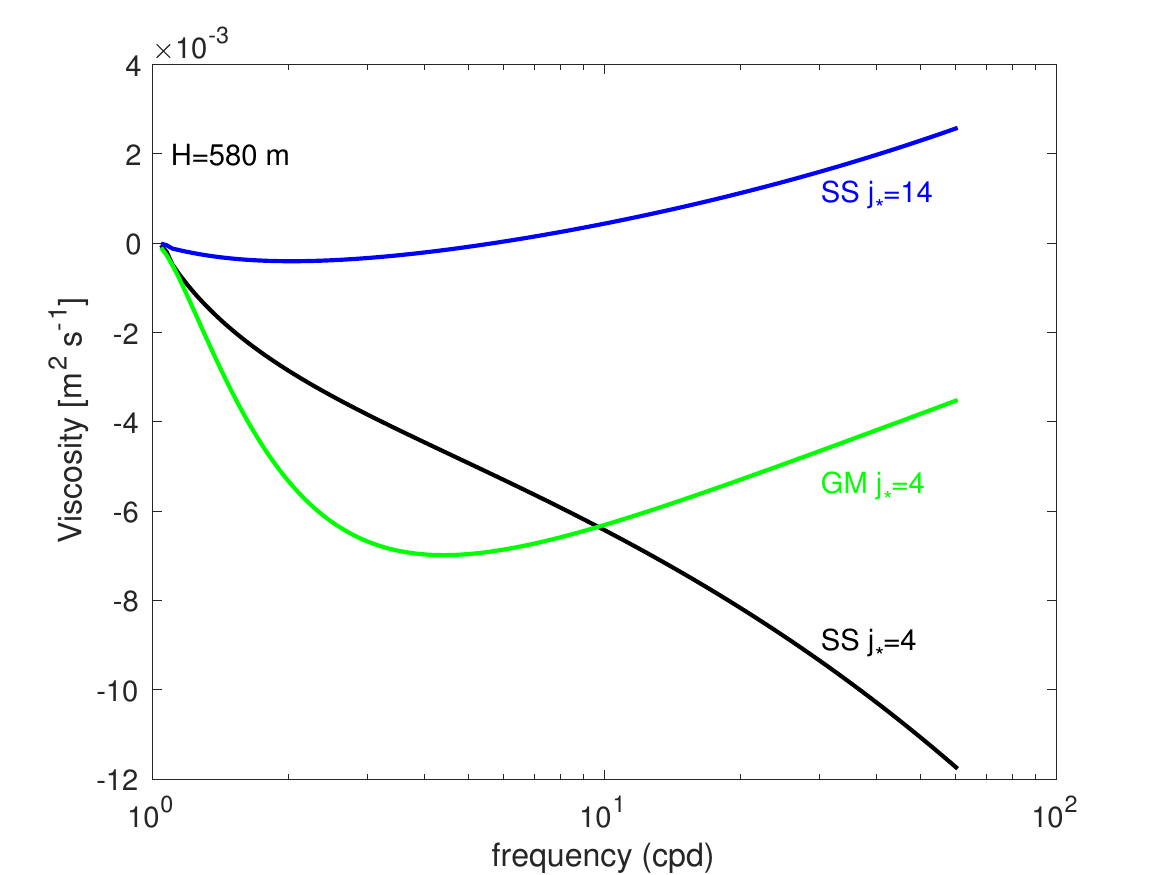}
\caption[]{Model estimates of the vertical viscosity for an eddy height scale of 1830/$\pi$ m.  The blue trace represents our reference Sargasso Sea spectrum, see figure \ref{diff_v} for the observational result, the black trace modifies that spectrum for an equivalent mode bandwidth of $j_{\ast}=4$, and the green trace represents the GM76 spectrum, determined by observations at Site-D, north of the GulfStream. 
}\label{model_vert_spectrum}
\end{figure}

\clearpage
\newpage








%



\bibliographystyle{ametsocV6}

\end{document}